\documentclass[aps,prd,10pt,nofootinbib,twocolumn,superscriptaddress,preprintnumbers]{revtex4-1}
\usepackage[english]{babel}
\usepackage[utf8]{inputenc}
\usepackage{latexsym} 
\usepackage{amsmath} 
\usepackage{amsfonts} 
\usepackage{amssymb} 
\usepackage{textcomp} 
\usepackage{graphicx}  
\usepackage{enumerate} 
\usepackage{array} 
\usepackage{mathrsfs}
\usepackage{amsthm}
\usepackage{bm,bbm}
\usepackage{bbold}  
\usepackage{multirow}
\usepackage{fancyref}
\usepackage{hyperref}
\usepackage{color}
\newcommand{\Eq}[1]{(\ref{#1})}

\newcommand{\B}[1]{\Blue{#1}}

\newcommand{\be}{\begin{equation}}
\newcommand{\ee}{\end{equation}}
\newcommand{\ba}{\begin{eqnarray}}
\newcommand{\ea}{\end{eqnarray}}
\newcommand{\bs}{\begin{subequations}}
\newcommand{\es}{\end{subequations}}
\def\com{\color{magenta}}
\def\cob{\color{blue}}
\newcommand{\rmd}{{\rm d}}
\newcommand{\rmi}{{\rm i}}
\newcommand{\rme}{{\rm e}}

\def\B{\Box}
\newcommand{\arX}[1]{\href{http://arxiv.org/abs/#1}{{\ttfamily\com arXiv:#1}}}
\newcommand{\oarX}[1]{\href{http://arxiv.org/abs/#1}{{\ttfamily\com arXiv:#1}}}

\newcommand{\procsin}[5]{in \emph{#1}, edited by #2 (#3, #4, #5)} 
\newcommand{\doin}[6]{\href{http://dx.doi.org/#1}{\cob #2\ #3 {\bf #4}, #5 (#6)}}

\newcommand{\doinn}[5]{\href{http://dx.doi.org/#1}{{\cob #2 {\bf #3}, #4 (#5)}}}
\newcommand{\doij}[5]{\href{http://dx.doi.org/#1}{{\cob #2 {#3} (#5) #4}}}

\newcommand{\tia}[1]{#1,}

\def\ra{\rightarrow}

\def\ove{\overline}

\def\p{\partial}
\def\a{\alpha}
\def\b{\beta}

\def\g{\gamma}
\def\s{\sigma}

\def\l{\lambda}
\def\n{\nu}
\def\m{\mu}

\def\om{\omega}

\def\vr{\varrho}

\def\e{\eta}
\def\d{\delta}

\def\ep{\epsilon}
\def\cA{\mathcal{A}}

\def\cD{\mathcal{D}}

\def\cK{\mathcal{K}}
\def\cL{\mathcal{L}}
\def\cP{\mathcal{P}}

\def\cF{\mathcal{F}}

\def\bl{\boldsymbol}
\renewcommand{\dh}{d_\textsc{h}}
\newcommand{\ds}{d_\textsc{s}}
\newcommand{\Pl}{{\text{\tiny Pl}}}
\newcommand{\mhyp}{\mbox{--}}

\begin{document}

\title{Standard Model in multiscale theories and observational constraints}
\author{Gianluca Calcagni}
\affiliation{Instituto de Estructura de la Materia, IEM-CSIC, Serrano 121, 28006 Madrid, Spain}
\email{calcagni@iem.cfmac.csic.es}
\author{Giuseppe Nardelli}
\affiliation{Dipartamento di Matematica e Fisica, Università Cattolica del Sacro Cuore, via Musei 41, 25121 Brescia, Italy}
\affiliation{TIFPA---INFN, c/o  Dipartimento di Fisica, Universit\`a di Trento, 38123 Povo (TN), Italy}
\email{giuseppe.nardelli@unicatt.it}
\author{David Rodr\'{\i}guez-Fern\'andez}
\affiliation{Departamento de F\'isica, Universidad de Oviedo, Avda.~Calvo Sotelo 18, 33007, Oviedo, Spain}
\affiliation{Departamento de F\'{\i}sica Te\'orica II, Universidad Complutense de Madrid, Parque de las Ciencias 1, 28040 Madrid, Spain}
\email{rodriguezferdavid@uniovi.es}

\begin{abstract}
We construct and analyze the Standard Model of electroweak and strong interactions in multiscale spacetimes with (i) weighted derivatives and (ii) $q$-derivatives. Both theories can be formulated in two different frames, called fractional and integer picture. By definition, the fractional picture is where physical predictions should be made. (i) In the theory with weighted derivatives, it is shown that gauge invariance and the requirement of having constant masses in all reference frames make the Standard Model in the integer picture indistinguishable from the ordinary one. Experiments involving only weak and strong forces are insensitive to a change of spacetime dimensionality also in the fractional picture, and only the electromagnetic and gravitational sectors can break the degeneracy. For the simplest multiscale measures with only one characteristic time, length and energy scale $t_*$, $\ell_*$ and $E_*$, we compute the Lamb shift in the hydrogen atom and constrain the multiscale correction to the ordinary result, getting the absolute upper bound $t_*<10^{-23}\,{\rm s}$. For the natural choice $\alpha_0=1/2$ of the fractional exponent in the measure, this bound is strengthened to $t_*<10^{-29}\,{\rm s}$, corresponding to $\ell_*<10^{-20}\,{\rm m}$ and $E_*>28\,{\rm TeV}$. Stronger bounds are obtained from the measurement of the fine-structure constant. (ii) In the theory with $q$-derivatives, considering the muon decay rate and the Lamb shift in light atoms, we obtain the independent absolute upper bounds $t_* < 10^{-13}{\rm s}$ and $E_*>35\,\text{MeV}$. For $\alpha_0=1/2$, the Lamb shift alone yields $t_*<10^{-27}\,{\rm s}$, $\ell_*<10^{-19}\,{\rm m}$ and $E_*>450\,\text{GeV}$.
\end{abstract}

\date{December 18, 2015}


\preprint{\doin{10.1103/PhysRevD.94.045018}{PHYSICAL REVIEW}{D}{94}{045018}{2016} \hspace{9cm} \arX{1512.06858}}


\maketitle


\section{Introduction and summary\\ of the main results}\label{intro}


\subsection{Dimensional flow and multiscale theories}

General relativity in four dimensions is an excellent description of spacetime and matter at low energies and large scales. However, as soon as gravity goes quantum, the very concept of smooth continuous geometry can break down at microscopic scales in favor of more abstract but more fundamental degrees of freedom. Fortunately, whenever this happens it is possible to find approximations such that the geometry retains at least some of its characteristics, \emph{in primis} the concept of spacetime dimension. Then, usually one is able to track the behavior of these features down to ultraviolet scales, for instance through the study of the spectral dimension $\ds$ and the Hausdorff dimension $\dh$. It is found that, in virtually all known approaches to quantum gravity, including string theory \cite{CaMo1}, either $\ds$ or $\dh$ (or both) run from 4 in the infrared to some value $\leq 2$ in the ultraviolet (see, e.g., \cite{tH93,Car09,fra1}). The transition between the two regimes varies depending on the model but it is continuous in general. Examples include causal dynamical triangulations \cite{AJL4,BeH,SVW1}, asymptotically safe quantum gravity \cite{LaR5,CES}, loop quantum gravity and spin foams \cite{Mod08,COT2,COT3}, Ho\v{r}ava--Lifshitz gravity \cite{SVW1,CES,Hor3}, noncommutative geometry \cite{Con06,CCM,AA} and $\kappa$-Minkowski spacetime \cite{Ben08,ArTr1}, nonlocal quantum gravity \cite{Mod11}, Stelle's gravity \cite{CMNa}, spacetimes with black holes \cite{CaG,Mur12,AC1}, fuzzy spacetimes \cite{MoN}, random combs \cite{DJW1,AGW}, random multigraphs \cite{GWZ1,GWZ2}, and causal sets \cite{EiMi}. 

In this context, we focus on theories of multiscale spacetimes \cite{fra1,fra4,frc1,frc2,ACOS,frc3,frc4,frc5,fra7,frc6,frc7,frc8,frc9,frc10,frc11,frc12}. These have been proposed either as stand-alone models of exotic geometry \cite{frc1,frc2,frc8,frc11} or as an effective means to study, in a controlled manner, the change of dimensionality with the probed scale (known as dimensional flow\footnote{We refrain from using the more common name ``dimensional reduction'' because, by a long-standing tradition, it indicates a completely different concept in Kaluza--Klein models, supergravity and string theory.}) in certain regimes of other quantum-gravity theories \cite{ACOS,frc4,fra7}. There exist four inequivalent multiscale theories of ``multifractional'' type which differ in the symmetries one imposes on the fundamental Lagrangian (Refs.\ \cite{frc11,AIP} provide pedagogical overviews of these models and of their status). The model with ordinary derivatives was the first to be proposed \cite{fra1,fra2,fra3} but it cannot be a fundamental theory due to some issues regarding its momentum space and quantization. The theory with fractional derivatives is most promising especially as far as renormalization is concerned but, apart from a general power-counting argument \cite{frc2}, its physical properties have not been studied yet; this will be done in the near future. The scenario with \emph{weighted derivatives} and the one with \emph{$q$-derivatives}, on both of which we focus here, have been analyzed in greater detail both in quantum field theory \cite{frc2,frc6,frc8,frc9} and cosmology \cite{frc11} and a wealth of new phenomenology has begun to emerge.

The motivation to consider these theories is threefold.
\begin{itemize}
\item[(A)] Multiscale spacetimes were originally proposed, with quantum gravity in mind, as a class of theories where the renormalization properties of perturbative quantum field theory could be improved, including in the gravity sector \cite{fra1,frc2}. Later on, it was shown that the two theories considered in the present paper do not have improved renormalizability \cite{frc9}, while the arguments of \cite{frc2} still apply to the theory with fractional derivatives. This specific incarnation of the multiscale paradigm is likely to fulfill the original expectations but, as we said, its study involves a number of technical challenges. However, massive evidence has been collected (and will be further increased by the findings of this paper) that \emph{all multifractional models share very similar properties} (e.g., \cite{frc2,frc4,frc10}). In preparation of dealing with the theory with fractional derivatives and to orient future research on the subject, it is important to understand in the simplest cases what type of phenomenology one has on a fractal spacetime. The theories with weighted and $q$-derivatives are simple enough to allow for a fully analytic treatment of the physical observables, while having all the features of multiscale geometries. Therefore, they are the ideal testing ground for these explorations. A better knowledge about the typical phenomenology occurring in multiscale spacetimes will be of great guidance for the study of the case with fractional derivatives.
\item[(B)] The theories with weighted and $q$-derivatives have not yet reached a satisfactory level of maturity and much needs to be done to assess their relevance and viability as physical models beyond the standard lore. For instance, the Standard Model has been formulated only in the electromagnetic sector \cite{frc8,frc10}, models of cosmic inflation and late-time acceleration have been explored only preliminarily \cite{frc11} and observational constraints on the fundamental scales of the geometry are either heuristic \cite{frc2} (based qualitatively on toy models of dimensional regularization) or too weak \cite{frc8}. Moreover, it is not even clear whether a satisfactory perturbative quantum field theory can be formulated at all in the case with weighted derivatives, due to difficulties in defining a predictive Feynman series of scattering processes \cite{frc9}. These theoretical problems deserve our attention.
\item[(C)] The search for experimental constraints on fractal spacetimes dates back to the 1980s \cite{ScM,ZS,MuS}. Since early proposals of fractal spacetimes were quite difficult to handle \cite{Sti77,Svo87,Ey89a,Ey89b}, toy models of dimensional regularization were used in \cite{ScM,ZS,MuS} and several bounds on the deviation of the spacetime dimension from 4 were obtained. However, these toy models were not backed by any specific framework and they were not even multiscale. Multifractional theories are a foundational proposal based on a complex combination of physical requirements with the principles of fractal geometry \cite{frc2}, they are multiscale and they provide a most natural and solid follow-up to the 1980s scenarios. The questions left unanswered by those models can now receive proper attention.
\end{itemize}


\subsection{Goals and results}\label{goa}

The first purpose of this paper is to construct the $SU(3)\otimes SU(2)\otimes U(1)$ Standard Model of electroweak and strong interactions in the multiscale theories with weighted and $q$-derivatives. We will extend the discussion started in \cite{frc8} for Abelian gauge fields and electrodynamics to non-Abelian fields.

The second goal is to see whether the problems found in \cite{frc9} for the case with weighted derivatives can be overcome and the theory made predictive. We answer in the affirmative. Gauge invariance and the requirement that all measurable masses are constant in all reference frames (a minimal requisite for a manageable perturbation theory) constrain the Lagrangian in such a way that field redefinitions from the so-called \emph{fractional picture} to the \emph{integer picture} (two inequivalent frames) map the model to the standard one. On one hand, this result goes against the general expectation that nonlinear interactions make such mapping impossible \cite{frc9}, thus avoiding the troubles that nonconstant couplings introduce in a quantum field theory. On the other hand, the ``trivialization'' of the model in the integer picture indicates that quantum fields may be insensitive to the anomalous properties of spacetime, in the absence of gravity. This is indeed the case for weak and strong interactions but not for quantum electrodynamics (QED); the reason of this discrepancy is that, \emph{among the gauge couplings, only the electric charge is observed directly}. The theory is nontrivial also because, when gravity is switched on, the field redefinitions are associated with a change of frame which never leads to general relativity plus minimally-coupled matter \cite{frc11}. Moreover, once the frame has been fixed, the measure affects the physics at mesoscopic scales, such as in thermodynamical and atomic systems \cite{frc14}. To summarize, we show that spacetimes with weighted derivatives are viable embeddings for a quantum field theory but their physical implications should be studied mainly in electrodynamics or at mesoscopic and large scales, especially in an astrophysical or cosmological setting.

The third goal of the paper is to extract, for the first time, physical predictions from the multiscale Standard Models. Since the weak and strong sectors with weighted derivatives are indistinguishable from the ordinary case, the question pertains only to electrodynamics for this theory, while in the theory with $q$-derivatives we have more nontrivial phenomenology at our disposal. The general strategy, originally embraced in early toy models of dimensional regularization \cite{ScM,ZS,MuS}, will be to use the experimental uncertainty of the most recent measurements of physical observables as an upper bound on the largest possible effect of multiscale geometry. This will allow us to place constraints on the time and length scales $t_*$ and $\ell_*$ below which geometry shows signs of a multifractal hierarchy.

In the theory with weighted derivatives, we consider the Lamb-shift effect in hydrogenic atoms and find the absolute upper bound
\be\label{tbou1}
t_* < 10^{-23}\,{\rm s}\,,
\ee
with a preferred range (i.e., for the natural choice $\a_0=1/2$ of the fractional exponent in the time direction, one of the parameters of the model)
\be\label{tbou3}
t_*^{(\alpha_0=1/2)} < 10^{-29}\,{\rm s}\,,
\ee
corresponding to energies $E_*>28\,{\rm TeV}$ right above the Large Hadron Collider (LHC) scale of $13\,{\rm TeV}$. 

The time scale $t_*$ can also be interpreted as the end of the era since the big bang (at $t=0$) when the universe showed a multiscale geometry. In this sense, the upper bound \Eq{tbou3} can be compared with the only other extant constraint
\be\label{oldco}
t_*^{(\alpha_0=1/2)}<10^6\,{\rm s}\approx 21\,{\rm days}\,,
\ee
obtained in \cite{frc8} from the variation $\Delta\a_\textsc{qed}/\a_\textsc{qed}$ of the fine structure constant at cosmological scales.\footnote{In \cite{frc8}, the expression $\Delta\a_\textsc{qed}/\a_\textsc{qed}\simeq -(1+\sqrt{t/t_*})^{-1}$ was found for a measure \Eq{mt2} with $\a_0=1/2$. Inverting with respect to $t_*$ and plugging in the mean value $\Delta\a_\textsc{qed}/\a_\textsc{qed}= (-0.57 \pm 0.11)\times 10^{-5}$ measured at Keck \cite{MWF,Mur03} and the age $t\approx 1.79\,{\rm Gyr}$ of the quasar, an estimate for $t_*$ was obtained. Taking instead the mean value as a rough upper bound on $\Delta\a_\textsc{qed}/\a_\textsc{qed}$, this estimate becomes the upper bound \Eq{oldco} for $t_*$.} The comparison illustrates a dramatic advancement. While the bound \Eq{oldco} is much weaker than what one should expect for consistency with the big-bang nucleosynthesis ($t_*<0.3\,{\rm s}$ \cite{frc8}), the constraint \Eq{tbou3} respects the nucleosynthesis bound and reduces the one from the fine-structure constant by 35 orders of magnitude! Another clear advantage of \Eq{tbou3} with respect to \Eq{oldco} is that it is based on well-established experimental determinations of the spectral lines of hydrogenlike atoms, while \Eq{oldco} is, at best, a heuristic estimate from observations that are still under debate. We therefore regard \Eq{tbou1} and \Eq{tbou3} as the first solid constraints on the time scale of the multiscale theory with weighted derivatives. Further bounds involving the fine-structure constant will be discussed at the end of the paper.

In the theory with $q$-derivatives, we consider also the muon decay rate, which gives independent information from the weak sector. One obtains
\be\label{tbou}
t_* < 10^{-13}{\rm s}\,,\qquad t_*^{(\alpha_0=1/2)} < 5\times 10^{-18}{\rm s}\,,
\ee
the first constraints ever on this theory. We will also find a lower bound on the fundamental energy scale $E_*$ in momentum space from the Lamb shift in the hydrogen atom,
\be\label{Elow}
E_*> 35\,\text{MeV}\,,\qquad E_*^{(\alpha_0=1/2)}> 450\,\text{GeV}\,.
\ee
At the end of the paper in Sec.\ \ref{sec:conc}, we will convert these bounds to stronger constraints on the time scale, $t_*<10^{-23}\,{\rm s}$ and $t_*^{(\alpha_0=1/2)}<10^{-27}\,{\rm s}$. 
 These numbers can be used in realistic cosmological models of the early universe to construct and test multiscale inflationary phenomenology. We will not pursue this line of investigation here. The bounds \Eq{tbou}--\Eq{Elow} have been announced in a companion paper \cite{frc12}. This theory avoids $\a_\textsc{qed}$ constraints.


\subsection{Comparison with more standard phenomenology beyond the Standard Model}

To explain our results intuitively to a bigger circle of phenomenologists, it might be useful to make a link with more familiar formulations of physics beyond the Standard Model. The presence of an underlying multiscale geometry affects the field theory in such a way that interaction terms (in gauge derivatives or in nonlinear potentials) acquire an explicit spacetime dependence of the form
\be\label{fff}
[1+f(x)]\phi_i\phi_j\cdots\,,
\ee
where $f(x)=f[v(x)]$ depends on the multiscale measure weight $v(x)$ characterizing the geometry and $\phi_i$ are some generic fields. Terms such as \Eq{fff} have a naive interpretation of ``having promoted coupling constants to fields'' and, in some sense, some of the physical effects we encounter are similar to those in models with varying couplings \cite{frc8}. Another possibility to mimic effects of the form \Eq{fff} might be to add higher-dimensional operators to a perfectly traditional Lagrangian. For instance, in a cubic single-scalar-field theory one would have $\phi^3\to[1+f(x)]\phi^3\sim [1+\phi^m+\phi^n+\cdots]\phi^3$ for some exponents $m$ and $n$, and one would fall into the context of effective field theories.

These are only superficial analogies which will be progressively screened out by various technical and conceptual \emph{distinguo} we will make in the paper. The most important of all is that $v(x)$ is not a scalar field and none of the above interpretations based on ordinary field theories has any such premade, nontrivial integrodifferential structure. Since $v(x)$ [hence $f(x)$] is fixed by the geometry, it cannot be interpreted as a field and the higher-order-operators comparison dies as soon as one writes down the classical or quantum dynamics [classically, one does not vary with respect to $v(x)$; at the quantum level, $v(x)$ does not propagate].\footnote{In scenarios with curved backgrounds, nondynamical scalars can still account for a nontrivial determinant of the metric. This is an incarnation of unimodular gravity that was compared with multiscale spacetimes in \cite[Sec.\ 3.1]{frc11}.} The varying-coupling interpretation can still survive but it loses any utility in the long run, since it does not explain why only certain couplings, but not others, depend on spacetime.


\subsection{Outline}

In Sec.\ \ref{sec:multi}, we briefly introduce the basic ingredients of multiscale spacetimes, giving much space to novel discussion on predictivity and falsifiability (Sec.\ \ref{predi}) and on the change of frame (Sec.\ \ref{sec:pic}).

The Standard Model of electroweak and strong interactions with weighted derivatives is constructed in Sec.\ \ref{ewwd}. In Sec.\ \ref{forma}, we set the formalism of Yang--Mills theory (Sec.\ \ref{sec:mills}) and interacting spinorial fields (Sec.\ \ref{spino}). 
 The Lagrangian of the electroweak model with weighted derivatives is constructed in Secs.\ \ref{sec:elec} and \ref{sec:lep}. The symmetries of the theory are discussed in Sec.\ \ref{simw}.

Section \ref{sec:vs} explores several conceptual features of the model with weighted derivatives of relevance for theory and experiments. The differences between the fractional and integer pictures, inequivalent frames related by field redefinitions, and reasons why one would not expect to have a quantum field theory under control are spelled out in Sec.\ \ref{sec:qft}. These issues are discussed in Sec.\ \ref{sec:fvi} in the case of the Standard Model: the theory is well defined but it does not give rise to characteristic predictions in accelerator experiments, apart in the electroweak sector (Sec.\ \ref{versus}). Possible deviations of the spacetime dimensionality from 4 are calculated in Sec.\ \ref{lashi}, where we place the bounds \Eq{tbou1} and \Eq{tbou3} on multiscale effects from the Lamb shift.

Section \ref{stmoq} is devoted to the theory with $q$-derivatives. The Standard Model on such spacetimes and its symmetries are introduced in Sec.\ \ref{qumu}, while in Sec.\ \ref{physq} we estimate how the anomalous geometry affects the muon lifetime $\tau_{\rm mu}$ and the Lamb shift. In Sec.\ \ref{sec:feynman}, we compute the correction $\Delta\tau=\tau_{\rm mu}-\tau_0$ to the standard value $\tau_0$ and extract the upper bound \Eq{tbou} on the characteristic time scale $t_*$ of the multiscale measure. A similar way to proceed is adopted in Sec.\ \ref{sec:lamb} for the  quantum electrodynamics corrections to the energy levels of light atoms, eventually leading to \Eq{Elow}. 

A discussion on further bounds on all the scales of both theories and conclusions are in Sec.\ \ref{sec:conc}.


\section{Review of multiscale spacetimes}\label{sec:multi}

We limit our attention to multiscale spacetimes defined on the ambient manifold $M_D$, $D$-dimensional Minkowski spacetime. The impact of curved backgrounds on this class of theories \cite{frc11} will be examined in Sec.\ \ref{withg}.


\subsection{Measure}\label{sec:mea}

The usual volume element $\rmd^Dx$ is replaced everywhere by the Lebesgue--Stieltjes measure $\rmd^Dx\ra \rmd\vr(x)$. In order to manipulate the measure, it is necessary to make some assumptions. (a) The measure is written as the standard Lebesgue measure times a non-negative weight factor,
\be \label{ntm}
\rmd\vr(x)=\rmd^Dx\,v(x)\,.
\ee 
(b) The weight $v(x)>0$ is a fixed coordinate profile where coordinates are factorized, $v(x)= \prod_\m v_\mu(x^\m)$, where the $D$ functions $v_\mu$ can be different from one another. (c) $v_\mu(x^\mu)=v_\mu(\vert x^\mu\vert)$.
 
The choice of weight is part of the definition of multiscale spacetimes. The goal is to define a measure on a continuum which could reproduce dimensional flow in quantum gravity or, more specifically and under certain approximations, a geometry with multifractal properties. This objective can be achieved by a specific set of rules which do not leave much liberty to the form of $v(x)$ \cite{frc1,frc2}. The simplest measure that one may use to obtain a noninteger dimensionality of spacetime is of the form
\be \label{smea}
v(x)=\prod_{\mu}{v_{\a_\mu}(x^\mu )}=\prod_{\mu=0}^{D-1}{\frac{{\vert x^\mu \vert}^{\alpha_\mu -1}}{\Gamma(\alpha_\mu)}}\,,
\ee
where $0<\a_\mu\leq 1$ are real-valued constants and $\Gamma$ is Euler's gamma function. It can be readily seen that, since $v$ has no dependence on any sort of characteristic scale, the measure weight \Eq{smea} leads to a constant Hausdorff dimension (the way a ball volume scales with its radius) $\dh= \sum_\mu\a_\mu\neq D$ rather than to a varying dimension.\footnote{We will keep referring to the Hausdorff dimension but similar considerations apply to the spectral dimension $\ds$ as well; see \cite{frc7} for details.} A more realistic \emph{Ansatz} is
\be \label{genmea}
v(x)=v_*(x):=\prod_\m\left[\sum^N_{n=1} g_{\m,n}(l^\m_n)v_{\a_n}(|x^\m|)\right],
\ee
where $N$ is integer and $g_{\m,n}$ are dimensionful coupling parameters which depend on the values of the characteristic length scales $l^\m_n$. To obtain dimensional flow, it is sufficient to consider a \emph{binomial} measure, $N=2$. In particular, to get $\dh=D$ in the infrared, we choose $g_{i,1}=\ell^{1-\a}_*$, $\a_i=\a$ for all spatial directions $i$, $g_{0,1}=|t_*|^{1-\a_0}$ and $g_{\m,2}=1=\a_{\m,2}$ for all $\mu$, where $\ell_*$ and $t_*$ are, respectively, a characteristic length and time. For instance, in $D=4$ topological dimensions we will consider the measure weight
\bs\label{binomialm}\ba
v_*({\bf{x}}) &=& \prod^{3}_{i=1}\left(1+\Bigg\vert\frac{x^i}{\ell_*}\Bigg\vert^{\a_i-1}\right),\\ 
v_*(t) &=& 1+\left|\frac{t}{t_*}\right|^{\a_0-1}.\label{mt2}
\ea\es
This is the simplest scale-dependent measure encoding a varying dimension. In the infrared and at late times ($x^i\gg \ell_*$, $t\gg t_*$), the Hausdorff dimension of spacetime is $\dh\simeq 4$, while in the ultraviolet and at early times $\dh\simeq 3\a+\a_0$. The transition between these regimes is smooth. 

Other measures more general than \Eq{genmea} are not only possible but also necessary if one wants to consider a geometry resembling a deterministic multifractal \cite{fra4,frc2}. In the simplest case (only one frequency $\om$), these measures are of the form \Eq{genmea} with the replacement $v_{\a_n}(x^\m)\to v_{\a_n}(x^\m) F_\om(\ln|x^\mu|)$, where for each direction (index $\mu$ omitted)
\be\label{fom}
F_\om(\ln|x|)=A\,\cos\left(\om\ln\left|\frac{x}{\ell_\infty}\right|\right)+B\,\sin\left(\om\ln\left|\frac{x}{\ell_\infty}\right|\right).
\ee
This modulation factor includes logarithmic oscillations and a fundamental scale $\ell_\infty$ much smaller than $\ell_*$, possibly of order of the Planck scale \cite{ACOS}. We will not use log-oscillating measures in the bulk of this paper, as the multifractional binomial measure \Eq{binomialm} will suffice for our purpose. However, we will invoke the scale $\ell_\infty$ in the conclusions to elaborate the constraints found from experiments.

The existence of a unitary invertible Fourier transform implies that also the measure in momentum space is anomalous,
\be\label{momes}
\rmd^Dk\to\rmd^Dp(k)=\rmd^Dk\, w(k)\,,
\ee
where $p^\mu(k^\mu)$ are geometric coordinates in momentum space, the weight $w(k)$ is factorizable and its form depends on the theory.

\subsubsection{Presentation, predictivity, and falsifiability}\label{predi}

Before moving on, several caveats deserve our attention. First, $v(x)$ \emph{is not a scalar field} but a distribution profile dictated by multifractal geometry. Therefore, it is not constrained dynamically. This is important not only because dynamics itself is strongly affected by the shape of $v(x)$ \cite{frc11}, but also as a means to tell apart our proposal from other Standard Models with varying couplings \cite{frc8,KiM}. 

Second, ordinary Poincar\'e invariance is violated by \Eq{genmea} and its concrete incarnation \Eq{binomialm}. This is a necessary price to pay to have a well-defined integrodifferential calculus: measures which preserve part of Poincar\'e invariance, for instance rotations, turn out to fare much worse than factorizable ones \cite{frc1}. However, the problem now arises of which coordinate frame one should choose to compare the theory with experiments. For example, one might pick a different polynomial measure $v_*(x)\to v_*(x-\bar x)$ peaked at a point $\bar x\neq 0$, and define ``infrared'' and ``ultraviolet'' with respect to the distance of an event from the new origin $\bar x$. Even if the measure is singular at a specific point $\bar x$, the singularity is integrable and it does not affect the well-definiteness of observables. As explained in \cite{frc1,frc8,turtl}, this is an issue of presentation of the measure that does not change the properties of the geometry (provided all the other parameters are kept fixed) but that, nevertheless, can affect physical predictions. This guarantees that, once the measure is completely chosen, multiscale theories are fully \emph{predictive}. Moreover, different presentations correspond to different theories and their physical differences can, in principle, be measured in certain experiments with the required sensitivity \cite{turtl}. Thus, regardless of whether the experiments we can make today reach such sensitivity or not, multiscale theories are also \emph{falsifiable}. Also, there are so few presentation choices (four have been identified in \cite{turtl}) that one cannot go on forever tailoring models in the case one presentation after another were excluded by observations. 

Let us now discuss the topics of presentation, predictivity, and falsifiability in relation with the results of the present paper. First, we choose a particularly natural presentation called ``null'' \cite{turtl}, which fixes $\bar x^\mu=0$ for all $\mu$. In many contexts, the null presentation coincides with another presentation called initial-point. For instance, in particle-physics experiments, one regards the point $\bar t$ as the beginning of the observation (the moment $t-\bar t=0$ when, e.g., a certain collision occurs or a certain particle is created) and $t_*$ as the time, measured from $\bar t$, before which multiscale effects are important. Setting $\bar t=0$, for symmetry with the rest of the coordinates one also has $\bar x^i=0$. In a cosmological system, $\bar t$ would be the discriminator between ``early'' times $\Delta t=t-\bar t\lesssim t_*$ and ``late'' times $\Delta t\gg t_*$. Here $\Delta t$ represents the moment when a cosmological phenomenon takes place with respect to some special instant $\bar t$ in the history of the universe. In this case, and without loss of generality, one defines $\bar t=0$ as the big bang, which is a most unique point in the cosmic evolution (it seems that multiscale cosmological models are not singularity free, in general \cite{frc11}). Again for symmetry, $\bar x^i=0$. The null presentation of the measure makes physical observables well defined; concrete examples will be seen in Secs.~\ref{lashi}, \ref{sec:feynman}, and \ref{sec:lamb}. For further details on this and other measure presentations, see \cite{turtl}.

Although we fix the presentation, some of the other parameters of the measure are left free (for instance, $\a$), so that we have to limit the discussion to constraints, rather than getting actual predictions. Even if we consider a type of phenomenology that does not lead to extremely strong constraints on the scales of the geometry, other observations have the potential to push these constraints to the point where specific models can be ruled out. It is therefore important to begin an extensive survey of the phenomenology of multiscale theories, ranging from particle physics to astrophysics and cosmology.

In what follows, we will use a generic weight $v(x)$ without specifying its form except in Secs.\ \ref{lashi}, \ref{sec:feynman}, \ref{sec:lamb}, and \ref{sec:conc}.
In those sections, the form of the measure is chosen to be the binomial \Eq{binomialm}, with or without log oscillations. Inclusion or not of log oscillations is just a matter of what scales are probed by an experiment. If these scales are expected to be larger than the fundamental scale appearing in Eq.\ \Eq{fom} (as is the case of the Standard Model, characterized by masses much smaller than the Planck mass), then one can consistently average out these oscillations to a first approximation. The choice to have only one scale $\ell_*$ among the infinitely many possible ones in the most general multiscale measure \Eq{genmea} (written explicitly in \cite{frc4}) is not based on a criterion of subjective simplicity. (If it were so, one would be able to consider more general scale hierarchies whenever a model is excluded by experiments, which would make the theory virtually unpredictive). In fact, the binomial measure can be regarded as an approximation of Eq.\ \Eq{genmea} where one picks the \emph{largest} scale $\ell_*$ in the hierarchy. Whatever happens at smaller scales, no matter the number of transient regimes with different dimensionalities from $\ell_*$ down to Planck scales, from the point of view of a macroscopic observer the first transition to an anomalous geometry will occur near $\ell_*$. Experiments constrain just this scale, the end of the multiscale hierarchy. Therefore, the fact that only one scale is sufficient to produce dimensional flow acts as a sort of superselection rule about the number of scales we should consider when confronting multifractional theories with experiments. Independently of the detailed behavior of the measure at ultrasmall scales, for any given presentation the physical consequences at scales $\sim\ell_*$ are universal and the theory is back-predictive.


\subsection{Theory with weighted derivatives}

In multiscale flat spacetimes with weighted derivatives, one replaces the usual Laplace--Beltrami operator $\B = \e^{\m\n}\p_\m \p_\n$ with
\be \label{wder}
\cK_v:=\eta^{\mu\nu}\cD_\mu \cD_\nu\,,\qquad \cD_\mu:=\frac{1}{\sqrt{v(x)}}\p_\mu \left[\sqrt{v(x)}\,\cdot\,\right],
\ee
where $\eta$ is the usual Minkowski metric with signature $(-,+,+,+)$. This choice of derivatives allows the construction of a momentum space with an invertible transform \cite{frc3} and has the advantage, contrary to fractional operators, to have simple composition rules. Defining
\be\label{chD}
\check{\cD}_\m=\frac{1}{v(x)}\p_\m\left[v(x)\,\cdot \,\right]\,,
\ee
one has
\ba
\check{\cD}_\m(A\cD^\m B) &=& \cD_\m A\cD^\m B +A\cD_\m\cD^\m B\,,\label{help1}\\
\cD_\m (AB) &=& (\cD_\m A)B+A(\p_\m B)\nonumber\\
&=& (\p_\m A)B+A(\cD_\m B)\,.\label{help2}
\ea
If we integrate the left-hand side of Eq.\ \Eq{help1} over the hypervolume \Eq{ntm}, the factor $v$ is canceled by the prefactor $1/v$ in $\check{\cD}$ and, for smooth fields vanishing at infinity, one establishes that $\int \rmd^Dx\,v(x)\, \cD_\m A\cD^\m B=-\int \rmd^Dx\,v(x)\,A\cD_\m\cD^\m B$. Also, since $v(x)$ is not a field, for an arbitrary variation $\delta$ of a field $\phi^i(x)$ ($i$ is a generic tensorial or family index), one has $\delta\left[v(x)\phi^i(x)\right]=v(x)\,\delta\phi^i(x)$. Moreover, from $\cD_\m \cD_\n\phi^i(x)=[v(x)]^{-1/2}\p_\m\p_\n[\sqrt{v(x)}\phi^i(x)]$, we have $[\cD_\m,\cD_\n]\phi^i =0$.

When constructing a field theory with weighted derivatives on Minkowski spacetime, one defines the action by replacing $\rmd^Dx\ra \rmd^Dx\,v(x)$ and $\p_\mu\to\cD_\mu$ in the corresponding standard action $\cL[\p_x,\phi^i(x)]$ (whatever it is) of fields $\phi^i$:
\be \label{action}
S=\int^{+\infty}_{-\infty}\rmd^Dx\, v(x)\,\cL[\cD_x,\phi^i(x)]\,,
\ee
where $\cL$ is the Lagrangian density. Equation \Eq{ntm} leads to a breaking of the Poincaré symmetries. In electromagnetism, noninvariance under translations gives rise to a Noether current not conserved in the familiar sense, $\p_\m J^\m\neq 0$. Instead, what one finds is the ``deformed'' conservation law \cite{frc8} 
\be \label{neth}
{\cD}_\m J^\m =0\,.
\ee
For an electromagnetic current $J^\mu$ characterized by a charge density $J^0=\rho$ and a flux vector ${\bf J}$, Eq.\ \Eq{neth} leads to the (non)conservation equation $-\cD_t\rho+\cD_i J^i=0$. Further, if we define the electric charge as 
\be\label{ech}
Q(t):=\int \rmd{\bf x}\,v({\bf x})\,\rho(t,{\bf x})\,,
\ee
one finds $\cD_t Q\neq 0\neq \dot Q$. This property opens up the possibility of having a varying electron charge \cite{frc8}. In the present work, we examine the implications of the anomalous background geometry also for the weak and strong sectors.


\subsection{Theory with \texorpdfstring{$q$}{}-derivatives}

Spacetimes with $q$-derivatives are much easier than those with weighted derivatives since they are invariant under the so-called $q$-Poincaré symmetries. The measure \Eq{ntm} can be rewritten as
\ba
&&\rmd\vr(x)=\rmd^D q(x)=\rmd q^0(x^0)\dots \rmd q^{D-1}(x^{D-1})\,,\\
&&q^\mu(x^\mu):=\int^{x^\mu}\rmd {x'}^\mu\,v_\mu ({x'}^\mu)\,,\label{quvi}
\ea
and the profiles $q^\mu(x^\mu)$ are called geometric coordinates. By definition, any Lagrangian $\cL[\p_x,\phi^i(x)]$ is replaced by $\cL\{\p_{q(x)},\phi^i[q(x)]\}$. In practice, one can pick the system of interest (Einstein gravity, the Standard Model, and so on) and simply make the replacement
\be\label{xq}
x\to q(x)
\ee
everywhere. The theory is then invariant under the nonlinear transformations
\be\label{qlort} 
{q'}^\mu(x^\mu)=\Lambda_\nu^{\ \mu}q^\nu(x^\nu)+a^\mu\,,
\ee
where $a^\mu$ is a constant vector.

The step \Eq{xq} leads to a nontrivial theory because part of the definition of multiscale spacetimes is the specification of measurement units for the coordinates. Time and spatial coordinates scale as lengths (in $c=1$ units), $[t]=-1=[x^i]$, which set our clocks and rods. On the other hand, geometric coordinates have an \emph{anomalous scaling} with respect to these clocks and rods and they represent mathematically and physically inequivalent objects with respect to the system $\{x^\mu\}$. Let us explain this point in detail. For the binomial measure \Eq{binomialm}, the geometric coordinates are
\bs\label{binomialm2}\ba
q_*^i(x^i) &=& x^i+\ell_*\frac{{\rm sgn}(x^i)}{\a_i}\Bigg\vert\frac{x^i}{\ell_*}\Bigg\vert^{\a_i},\\ 
q_*(t) &=& t+t_*\frac{{\rm sgn}(t)}{\a_0}\left|\frac{t}{t_*}\right|^{\a_0}.
\ea\es
Although $[q^\mu(x^\mu)]=-1$ at all scales just like the coordinate $x^\mu$, at different regimes its $x$-dependence changes and, in the ultraviolet, one has $q\propto x^\a$ and an anomalous scaling for $\a\neq 1$. To see that the $q$-theory is inequivalent to the standard one, one can look at the structure of momentum space and at its consequences, for instance in the primordial cosmological spectra of inflation \cite{frc11}. The expression of the measure $\rmd^Dp(k)$ in momentum space and of its coordinates $p^\mu(k^\mu)$ is universal and independent of the form of the spacetime geometric coordinates:
\be\label{pik}
p^\mu(k^\mu)=\frac{1}{q^\mu(1/k^\mu)}\,,\qquad l_n^\mu \leftrightarrow \frac{1}{k_n^\mu}\,,
\ee
under the provision that the hierarchy of length scales $\{l_n^\mu\}$ appearing in $q^\mu$ be replaced by a hierarchy of energy-momentum scales $\{k_n^\mu\}$. A further simplification occurs if all momentum scales along different directions collapse to just one energy scale $E_{*n}$. For example, the momentum geometric coordinates \Eq{pik} associated with the binomial measure \Eq{binomialm2} are
\bs\label{binomialm2k}\ba
p_*(k^i) &=& \left[\frac{1}{k^i}+\frac{1}{E_*}\frac{{\rm sgn}(k^i)}{\a_i}\Bigg\vert\frac{E_*}{k^i}\Bigg\vert^{\a_i}\right]^{-1},\\ 
p_*(E) &=& \left[\frac{1}{E}+\frac{{\rm sgn}(E)}{E_*\a_0}\left|\frac{E_*}{E}\right|^{\a_0}\right]^{-1}.\label{bino2}
\ea\es

Due to its simplicity, the construction of the Standard Model in this class of spacetimes will take much less effort than for the theory with weighted derivatives.

 
\subsection{Pictures and physical observables}\label{sec:pic}

The structure of \Eq{wder} has suggested, since early stages, a convenient way to recast systems with weighted derivatives into a more familiar one. Given an action $S_\eta[v,\cD,\phi^i,m_i,\l_i]$ with integration measure weight $v(x)$, Minkowski metric $\eta_{\mu\nu}$, weighted derivatives $\cD_\mu$, matter fields $\phi^i$, masses $m_i$ and couplings $\lambda_i$, if the kinetic terms are at most quadratic one can make field redefinitions
\be\label{vpph}
\tilde\phi^i=\sqrt{v}\,\phi^i
\ee
such that the following mapping holds:
\be\label{fip}
S_\eta[v,\cD,\phi^i,m_i,\l_i]=\tilde S_\eta[1,\p,\tilde\phi^i,m_i,\tilde \l_i]\,,
\ee
where the couplings have been redefined accordingly and the masses remain the same ($v m_i^2\phi_i^2=m_i^2\tilde\phi_i^2$, see below). The left-hand side of \Eq{fip} is the starting point where the multiscale theory is defined and the anomalous geometry is manifest; the set of variables $\{\phi^i,m_i,\l_i\}$ is called \emph{fractional picture}. The right-hand side of \Eq{fip} looks like a field theory in ordinary spacetime, where $\a_\mu=1$; the set of variables $\{\tilde\phi^i,m_i,\tilde \l_i\}$ is then called \emph{integer picture}. 

The theory with $q$-derivatives is even simpler to formulate. There is no field redefinition analogous to \Eq{vpph} and the mapping \Eq{fip} is replaced by
\be\label{fipq}
S_\eta[v,v^{-1}\p_x,\phi^i,m_i,\l_i]=S_\eta[1,\p_q,\phi^i,m_i,\l_i]\,.
\ee
In this case, we will call fractional picture the frame where the $x$-dependence of the geometric coordinates $q(x)$ is manifest [left-hand side of \Eq{fipq}] and integer picture the frame described by the geometric coordinates $q$ [right-hand side of \Eq{fipq}].

The difference between the fractional and the integer picture is in the way geometry is perceived by the dynamical degrees of freedom: as standard Minkowski spacetime in the integer picture, as an anomalous geometry with a fixed integrodifferential structure in the fractional picture. The presence of this predetermined structure does affect the physics because it prescribes the existence of a preferred frame where physical observables should be compared with experiments. \emph{By definition} of the theory, this frame is the fractional picture. Even if some or all the steps of the calculation of such observables can (or, for quantum field theory, \emph{must}) be done in the integer picture, the final result must be reconverted back to the fractional picture. Roughly speaking, not doing so would amount to get wrong numbers from a set of adaptive $q$-clocks and $q$-rods \cite{turtl}. This is an important conceptual novelty with respect to theories with an ordinary integrodifferential structure: a choice of frame is a mandatory step in the definition of multiscale spacetimes. 

In the case with $q$-derivatives, time intervals, lengths and energies are physically measured in the fractional picture with coordinates $x^\mu$ ($k^\mu$ in momentum space), where coordinate transformations are described by the nonlinear law \Eq{qlort}. It may be useful to stress that Eqs.~\Eq{xq} and \Eq{quvi} are \emph{not} a coordinate transformation. They govern the passage between the fractional picture and the integer picture described by the composite coordinates $q(x)$ [$p(k)$ in momentum space].

The case with weighted derivatives is more delicate because the triviality of the right-hand side of the mapping \Eq{fip} depends on the system under consideration. Moreover, even in cases where the right-hand side is trivial (i.e., when $\tilde\lambda_i={\rm const}$), it is not obvious that physical observables will be trivial, too. Therefore, even if we have defined the theory to give predictions in the fractional frame, one should verify explicitly that these predictions are nontrivial. This point is better illustrated by concrete examples and, for this reason, we will postpone its discussion to Sec.\ \ref{sec:vs}. Here we anticipate the gist of it: the Standard Model will turn out to be trivial in the integer picture but the electrodynamics sector will, nevertheless, give rise to nontrivial observables in the fractional picture. The situation becomes much clearer in the presence of gravity because, in that case, the system can never be trivialized in the integer picture; see Sec.\ \ref{withg}.

Finally, we make a remark on the multiscale theories with ordinary and fractional derivatives, which we do not consider in this paper. Models with ordinary derivatives  
can at best be regarded as an effective description of anomalous spacetimes, since they suffer from several problems (see \cite{frc11} for a recapitulation). Still, whenever trustable, their predictions are nontrivial: actions have the form $S_g[v,\p,\phi^i,m_i,\l_i]$ in a generic embedding with metric $g_{\mu\nu}$ and there is no such thing as an integer picture. To the best of our knowledge, also the theory with fractional derivatives cannot be trivialized in a suitable frame, due to the complexity of the differential structure.


\section{Standard Model with\\ weighted derivatives}\label{ewwd}


\subsection{Gauge fields, fermions and varying couplings}\label{forma}

\subsubsection{Gauge transformations}\label{sec:mills}

In this subsection, the infinitesimal and finite gauge transformations for gauge fields and spinors are established.

We define the Yang--Mills field $\boldsymbol{\cA}_\m =A_\m\!^a \bl{t}_a$,\footnote{Hereafter, we will use Latin indices $a,b,\dots$ when we refer to inner degrees of freedom related to the generators of a Lie group, and Greek indices $\mu,\nu,\dots$ when we refer to spacetime coordinates. We work in $D$ topological dimensions until we compare the theory with experiments. Latin indices $i, j, \dots$ run over the specific representation of the group. If not specified, the fundamental representation will be employed, and $i,j,\ldots = 1, \dots, n$ for $SU(n)$. Also, we will use boldface fonts when internal indices are contracted and normal case when we refer to field components.} where $A_\m\!^a$ is a non-Abelian vector field and $\bl{t}^a$ are the matrix representations of the Lie algebra $[\bl{t}_a ,\bl{t}_b]=\rmi f_{abc}\bl{t}^c$, together with the normalization condition ${\rm tr}(\bl{t}^a\bl{t}^b)=\frac{1}{2}\d^{ab}$. In \cite{frc8}, the covariant derivative for the Abelian gauge group $U(1)$ was defined as (in $\hbar=1=c$ units)
\be \label{covderu1}
\nabla_\mu:= \cD_\mu + \rmi e A_\mu\,,
\ee
with $e$ the charge that couples to electromagnetism. For the multiscale theory with weighted derivatives, one has
\be \label{e}
e(x)=\sqrt{v(x)}e_0\,,
\ee
$e_0$ being the usual electron charge.\footnote{Unless otherwise specified, we shall use the subscript $0$ to denote constant couplings.}

Let us now consider the Lie group $SU(n)$ with an arbitrary $n$. Let $\Psi_i(x)$ be the components of a matter field 
transforming according to a given representation ${\bf t}_a$ of the Lie algebra. Making the symmetry local, the infinitesimal transformation reads
\be \label{transf}
\d \Psi_i=\rmi g\,\ep^c\,{(t_c)_i}^j\Psi_j\,,
\ee
 where $\ep^c=\ep^c(x)$ denote the components of a set of $n^2-1$ functions which depend on the coordinates and $g= g(x)$ is a charge which, in principle, may vary in space and time. We will keep the coordinate dependence of $\ep^c$ and $g$ implicit in what follows. Differentiating Eq.\ \Eq{transf} gives
\be \label{transfder}
\d(\cD_\m \Psi)_i=\rmi[\p_\mu(g\ep^c)\,(t_c)_i\!^j\Psi_j+ g\ep^c\,{(t_c)_i}^j\cD_\m\Psi_j]\,,
\ee
where we used Eq.\ \Eq{help2}. In order to make the kinetic term of the Lagrangian invariant under Eq.\ \Eq{transfder}, it is required that the derivative operator transforms just like $\Psi$ itself. To do so, we make the replacement
\be \label{gcov}
\cD_\m \ra \nabla_\m=\cD_\m +\rmi g\boldsymbol{\cA}_\m\,,
\ee
so that we have $\d (\nabla_\m \Psi)_i=\rmi g\ep^a\,{(t_a)_i}^j\nabla_\m\Psi_j$ if, and only if, the variation of $\boldsymbol{\cA}_\m$ is 
\ba
\left[\d (g\cA_\m\Psi)\right]_i &=& \rmi g^2\ep^a\,A_\m\!^b f^{\ \ c}_{ab}\,{(t_c)_i}^m\Psi_m\nonumber\\
& & -\p_\m ( g\ep^b)\,{(t_b)_i}^j\Psi_j\,,
\ea
for any matter field $\Psi$. The last expression can be rewritten as
\be \label{ainfi} 
\d\left(g \bl{\cA}_\m\right) =g^2\left[\rmi\boldsymbol{\ep},\bl{\cA}_\m\right]-\p_\m (g\boldsymbol{\ep})\,,
\ee
with $\boldsymbol{\ep}=\ep^a{\bf t}_a$. This is the infinitesimal transformation of the gauge field $\boldsymbol{\cA}_\m$. In electromagnetism, gauge invariance yields the relation \Eq{e} between $e$ and $e_0$.

Next, we define a finite gauge transformation as
\be \label{fpsi}
\Psi'(x)=\om(x)\Psi(x)\,,
\ee
or, in components, $\Psi_i'(x)=\om_i\!^j(x)\Psi_j(x)$. In this case, the derivative operator defined in Eq.\ \Eq{gcov} will be covariant if, and only if,
\ba \label{atrans}
\rmi g\cA'^a_\m\,{(t_a)_j}^k &=& \rmi g\,\om_j\!^l \cA^a_\m\,(t_a)_l\!^m(\om^{-1})_m\!^k-\p_\m\om_j\!^n(\om^{-1})_n\!^k\,,\nonumber\\
\ea
or, in a more compact form,
\be 
\rmi g\boldsymbol{\cA}'_\m=\rmi g\,\om\boldsymbol{\cA}_\m\om^{-1}-(\p_\m\om)\,\om^{-1}\,.
\ee
In particular, taking $U(1)$, $g=e$, $\bl{\cA}_\m =A_\m$, $\om =\rme^{\rmi e\l(x)}$, from Eq.\ \Eq{e} we find that $A'_\m$ and $A_\m$ are related by $A'_\m=A_\m +\cD_\m\l$, consistently with \cite{frc8}.

From here, we recall that the anomalous geometry does not modify the definition of the group $SU(n)$, which is still arc connected. Thus we can expand $\om$ in a neighborhood of the identity, so that $\boldsymbol{\om}\simeq \mathbbm{1}+\rmi g\boldsymbol{\ep}$ or, in components, $\om_i\!^j\simeq\d_i\!^j +\rmi g\ep^a(t_a)_i\!^j$. With this, it is straightforward to see that the infinitesimal transformation $\boldsymbol{\cA}'_\m -\boldsymbol{\cA}_\m =\d \boldsymbol{\cA}_\m$ is indeed given by Eq.\ \Eq{ainfi}.

Finally, we define the field strength or curvature tensor $\boldsymbol{\cF}_{\m\n}=\cF^a\!_{\m\n}{\bf t}_a$ as the commutator of the double covariant derivative \Eq{gcov}, acting on $\Psi$:
\be \label{F}
\rmi g\boldsymbol{\cF}_{\m\n}\Psi =\left[\nabla_\m ,\nabla_\n\right]\Psi\,.
\ee
Substituting Eq.\ \Eq{gcov} in \Eq{F}, we get
\ba \label{F2}
\cF_{\m\n}^a\,(t_a)_i\!^j\Psi_j &=& \frac{1}{g}\left[\p_\m(g\cA^b_\n)(t_b)_i\!^j
 -\p_\n(g\cA^c_\m)(t_c)_i\!^j\right]\Psi_j\nonumber\\
 & & -gA^c_\m A^d_\n f^e_{cd}\,(t_e)_i\!^j\Psi_j \,,
\ea
where we used Eq.\ \Eq{help2}. Combining Eqs.\ \Eq{atrans} and \Eq{F}, the transformation law for $\bl{\cF}_{\m\n}$ under finite gauge transformations follows: 
\be \label{finv}
\boldsymbol{\cF}'_{\m\n}(x)=\om(x)\,\boldsymbol{\cF}_{\m\n}(x)\,\om^{-1}(x)\,,
\ee
whereas under infinitesimal gauge transformations one has $\d\bl{\cF}_{\m\n}=\bl{\cF}'_{\m\n}-\bl{\cF}_{\m\n}=\rmi g\left[\bl{\ep},\bl{\cF}_{\m\n}\right]$ or, in components, $\cF^a_{\m\n}\ra \cF^a_{\m\n} -gf^a_{bc}\cF^c_{\m\n}\ep^b$, which is the same for a constant $\ep$ or a local $\ep(x)$.

\subsubsection{Interacting fermions}\label{spino}

In Eq.\ \Eq{e}, the $U(1)$ electric charge may vary in spacetime \cite{frc8,frc10}. In this subsection, we shall determine whether it is possible to obtain the same dependence for the group $SU(n)$. The physically relevant groups are $SU(2)$ and $SU(3)$.

Let us consider an interacting theory of a gauge field $\bl{\cA}_\m$ and a fermion field $\Psi$ invariant under the combined local gauge transformations \Eq{fpsi} and \Eq{finv}. By imposing uniqueness for the conservation law for the Noether current, we shall determine the relation between the value of the coupling \emph{constant} $g_0$ found in the standard theory and $g$. We will start from the gauge-invariant Lagrangian density 
\bs\label{lagyan}\ba
\cL&=&\cL_{\rm YM}+\cL_{\rm int} +\cL_m\,,\\
\cL_{\rm YM}&=&-\frac{1}{2}{\rm tr}(\bl{\cF}_{\m\n}\bl{\cF}^{\m\n})\,,\\
\cL_{\rm int} &=& \rmi\ove{\Psi}\g_\m\left(\cD^\m +\rmi g\bl{\cA}^\m\right)\Psi\,,\label{lint}\\
\cL_m &=& -m\ove{\Psi}\Psi\,,
\ea\es
where $\g^\m$ are the usual Dirac matrices, $\cL_{\rm int}$ is the contribution to the Lagrangian due to interaction between fields $\Psi$ and $\boldsymbol{\cA}_\m$, and $\cL_m$ is the mass term. All the contributions $\cL_{\rm YM}$, $\cL_{\rm int}$ and $\cL_m$ are gauge invariant separately. We now define
\bs\label{ca}\ba
g(x) =:\sqrt{v(x)}\,g_v(x)\,,\quad \tilde{\Psi}:=\sqrt{v(x)}\Psi\,,\\
 g_0\tilde{\bl{\cA}}_\m :=g(x)\bl{\cA}_\m=\sqrt{v(x)}\,g_v(x)\bl{\cA}_\m\,,
\ea\es
where $g_0$ is constant but $g_v$ varies in spacetime. Substituting \Eq{ca} into Eq.\ \Eq{F}, 
\ba
{\bl{\cF}}_{\m\n} &=& \frac{1}{\sqrt{v}}\frac{g_0}{g_v}\Big\lbrace\p_\m\tilde{\bl{\cA}}_\n -\p_\n\tilde{\bl{\cA}}_\m +\rmi g_0[\tilde{\bl{\cA}}_\m,\tilde{\bl{\cA}}_\n]\Big\rbrace\nonumber\\
&=& \frac{1}{\sqrt{v}}\frac{g_0}{g_v}\tilde{\bl{\cF}}_{\m\n}\,,\\
\nonumber
\ea
where $\tilde{\bl{\cF}}_{\m\n}:=\p_\m\tilde{\bl{\cA}}_\n -\p_\n\tilde{\bl{\cA}}_\m +{i}g_0[\tilde{\bl{\cA}}_\m,\tilde{\bl{\cA}}_\n]$. With this, the Lagrangian density \Eq{lagyan} reads
\ba
\hspace{-0.5cm}\cL &=&\frac{1}{v}\left[-\frac{1}{4}\frac{g_0^2}{g_v^2}(\tilde{\cF}^a_{\m\n}\tilde{\cF}^{\m\n}_a)+\rmi\ove{\tilde{\Psi}}\g_\m\tilde\nabla^\mu\tilde{\Psi} -m\ove{\tilde{\Psi}}\tilde{\Psi}\right],\label{tilL}\\
\hspace{-0.5cm}\tilde\nabla^\mu &=& \p^\m +\rmi g_0\tilde{\bl{\cA}}^\m\,,
\ea
where we raise and lower indices with the Minkowski metric. Now, the factor $1/v(x)$ annihilates the $v(x)$ that appears in Eq.\ \Eq{action}, so that the action functional is the same as usual, apart from the fact that the Yang--Mills coupling in front of the $\tilde{\cal F}^2$ term might be spacetime dependent (we will presently come back to this point). Hence, varying the action with respect to $\tilde{{\cA}}^a_\n$ yields the equations of motion
\be \label{cov1}
\p_\n\left(\frac{g_0^2}{g_v^2}\tilde{\cF}_a^{\m\n}\right)-\frac{g_0^3}{g_v^2}f_a^{\ bc}\tilde{\cA}_{b\n} \tilde{\cF}_c^{\m\n} =-g_0\ove{\tilde{\Psi}}\g^\m\bl{t}_a\tilde{\Psi}\,.
\ee

Applying $\p_\m$ to Eq.\ \Eq{cov1} and noting that $\p_\m\p_\n(g_0^2\tilde{\cF}_a^{\m\n}/g_v^2)=0$ since $\tilde\cF_a^{\m\n}$ is antisymmetric in the spacetime indices, one gets
\be\label{noether3}
0=\p_\m \tilde{j}_a^\m:=\p_\m\left(-g_0\ove{\tilde{\Psi}}\g^\m\bl{t}_a\tilde{\Psi}+ \frac{g_0^3}{g_v^2}f_a^{\ bc}\tilde{{\cA}}_{b\n}\tilde{\cF}_c^{\m\n}\right),
\ee
or, before the field redefinitions \Eq{ca}, the equations of motion \Eq{cov1} and Eq.\ \Eq{noether3} for the $n^2-1$ Noether currents read
\ba
\cD_\mu \cF^{\mu\nu} &=& j^\nu\,,\label{maxna}\\
\cD_\m j^\m_a &=& 0\,,\label{noether4}\\
j^\m_a &:=& -g\ove{{\Psi}}\g^\m\bl{t}_a{\Psi}+\frac{g_0}{g_v}gf_a^{\ bc} {\cA}_{b\n}{\cF}_c^{\m\n}.
\ea
Nonconservation is caused by the nontrivial weight factors, as one can check when integrating Eq.\ \Eq{noether4} over a hypervolume $\int \,\rmd^Dx\,v(x)$. Also, and as in the ordinary case $v=1$, $j^\m_a$ is not gauge invariant, due to the presence of the term $\propto {\cA}_\n{\cF}^{\m\n}$. In the Abelian case with weighted derivatives, Maxwell's equations are $\cD_\mu F^{\mu\nu} = J^\nu$ \cite{frc8}. The left-hand side is gauge invariant and, therefore, so is the right-hand side. Saturating with $\cD_\nu$ yields the deformed conservation law \Eq{neth}. In the non-Abelian case, the left-hand side of Eq.\ \Eq{maxna} belongs to the adjoint representation, implying that the right-hand side is not gauge invariant.

In response to this, one can define the matter current 
\be
J^\m_a:=-g\,\ove{\Psi}\g^\m \bl{t}_a\Psi\,,
\ee
which is gauge invariant and obeys the law
\ba \label{noether5}
\nabla_\m J^\m_a &=& \cD_\m J^\m_a -f_a^{\ bc}\cA_{b\m}J_c^\m \nonumber\\
&=& -\cD_\m\left(g\ove{\Psi}\g^\m \bl{t}_a\Psi\right)+g f_a^{\ bc}\cA_{b\m}\ove{\Psi}\g^\m \bl{t}_a\Psi\nonumber\\
&=& 0
\ea
in the adjoint representation. On the other hand, from the Euler--Lagrange equations
\be
\p_\m\left[\frac{\p\tilde{\cL}}{\p(\p_\m\tilde{{\Psi}}^a)}\right]=\frac{\p\tilde{\cL}}{\p\tilde{{\Psi}}^a}\,,\quad \p_\m\!\left[\frac{\p\tilde{\cL}}{\p\left(\p_\m\ove{\tilde{\Psi}}^a\right)}\right]=\frac{\p\tilde{\cL}}{\p\ove{\tilde{\Psi}}^a}\,,
\ee
the Dirac equation with electromagnetic interactions and its conjugate are
\ba
\rmi\g^\m\p_\m\tilde{\Psi}-m\tilde{\Psi}&=&g_0\tilde{\bl{\cA}}_\m\g^\m\tilde{\Psi}\,\label{cov2},\\
\rmi\p_\m\ove{\tilde{\Psi}}\g^\m +m\ove{\tilde{\Psi}}&=&-g_0\tilde{\bl{\cA}}_\m\ove{\tilde{\Psi}}\g^\m\,.\label{covx}
\ea
Multiplying \Eq{cov2} by $\tilde{\Psi}$, \Eq{covx} by $\ove{\tilde{\Psi}}$ and taking the sum, we find
\be \label{noether2}
\p_\m\left(\ove{\tilde{\Psi}}\g^\m\tilde{\Psi}\right)=0\qquad\Rightarrow\qquad \check{\cD}_\m\left(\ove{\Psi}\g^\m\Psi\right)=0\,,
\ee
where we used the weighted derivative \Eq{chD}. This is the Noether current arising from the symmetry under $U(1)$ transformations of the Lagrangian density \Eq{lagyan}.

Equation \Eq{noether5} allows us to set a relation between $g$, $g_v$ and $g_0$. By virtue of the $U(1)$ symmetry, both \Eq{noether4} and \Eq{noether5} must reduce to \Eq{noether2} when taking $f_{abc}=0$. This happens only if 
\be \label{g}
g_v=g_0\,,\qquad g(x)=\sqrt{v(x)}g_0\,.
\ee
These relations can also be obtained by noticing that $\tilde j^\mu_a = (g_0/g_v)\sqrt{v} j^\mu_a$ and requiring the usual vector-field transformation between the fractional and the integer picture.

Taking into account Eq.\ \Eq{g}, we recast \Eq{F2} as $\bl{\cF}_{\m\n}=2\cD_{[\m}\bl{\cA}_{\n]}+\rmi g\left[\bl{\cA}_\m,\bl{\cA}_\n\right]$ and the equations of motion \Eq{cov1} and \Eq{cov2} in the fractional picture:
\ba \label{meq}
\nabla_\n\boldsymbol{\cF}^{\m\n} &=&\cD_\n \boldsymbol{\cF}^{\m\n} +\rmi g[\boldsymbol{\cA}_\n,\bl{\cF}^{\m\n}]=-g\ove{\Psi}\g^\m\Psi\,,\nonumber\\
0&=& \rmi\g^\m\left(\cD_\m\Psi +\rmi g\bl{\cA}_\m\right)\Psi -m\Psi \,.
\ea

One may wonder whether spacetime-dependent couplings are a generic feature of multiscale models with weighted derivatives \cite{frc9}, and whether also masses acquire such dependence, as expected from simple considerations in special relativity \cite{frc10}. In this paper, we show that variable charges do not lead unavoidably to variable masses and that we can have a theory with varying charges but with constant masses. 


\subsection{Multiscale electroweak model: Bosonic sector}\label{sec:elec}

We proceed to study the electroweak sector of the Standard Model. Its fundamental degrees of freedom are massless spin-$1/2$ chiral particles and the gauge symmetry group is $SU(2)_{\rm L}\otimes U(1)$, where $SU(2)_{\rm L}$ acts only on left fermions and $U(1)$ is the weak-hypercharge symmetry. In the usual theory, the coupling constants are $g'_0$ and $g_0$, respectively, and they are related to the couplings in the fractional picture by \Eq{g}, $g=\sqrt{v}\,g_0$ and $g'=\sqrt{v}\,g_0'$.

Let us denote by $A_\m^a$ and $B_\m$ the gauge fields of $SU(2)$ and $U(1)$, respectively. The generators of $SU(2)$ are $\sigma_a/2$, $\sigma_a$ being the Pauli matrices. The gauge covariant derivative acting on a complex isodoublet $\Phi$ with hypercharge $Y=1/2$ is 
\ba
\nabla_\m\Phi &=&\left(\cD_\m +\frac{\rmi}{2} g'\sigma_a A_\m^a +\rmi g Y B_\m\,\right)\!\Phi\,,\nonumber\\
 &=& \left[\cD_\m +\frac{\rmi}{2} g'\left(\begin{matrix}
A_\m^3 & A_\m^1-\rmi A_\m^2 \\
A_\m^1 +\rmi A_\m^2 & -A_\m^3 \\
\end{matrix}\right) +\frac{\rmi}{2} gB_\m\right]\!\Phi.\nonumber\\\label{nab1}
\ea
The field-strength tensors are defined according to \Eq{F2} that, by taking \Eq{g} into account, reads
\ba
F^a_{\m\n} &=& \cD_\m A^a_\n -\cD_\n A^a_\m - g'\epsilon^a_{\ bc} A_\m^b A_\n^c\,,\label{fwei}\\
B_{\m\n} &=&\cD_\m B_\n -\cD_\n B_\m\,,\label{bwei}
\ea
where we used the structure constants of the $SU(2)$ gauge group, $f^{abc}=\epsilon^{abc}$. 

In the integer picture, from \Eq{vpph} the above covariant derivative can be written as $[v(x)]^{-1/2}\tilde{\nabla}_\m\tilde\phi^i(x)$ and
\ba
\tilde{\nabla}_\m &=& \left[\p_\m +\frac{\rmi}{2}g_0'\left(\begin{matrix}
	\tilde{A}_\m^3 & \tilde{A}_\m^1-\rmi\tilde{A}_\m^2 \\
	\tilde{A}_\m^1 +\rmi\tilde{A}_\m^2 & -\tilde{A}_\m^3 \\
\end{matrix}\right) +\frac{\rmi}{2}{g_0}\tilde{B}_\m\right],\nonumber\\
\ea 
while the field strengths are
\ba
\tilde{F}^a_{\m\n}&=&\p_\m \tilde{A}^a_\n -\p_\n \tilde{A}^a_\m -g'_0 \epsilon^{abc} \tilde{A}^b_\m \tilde{A}^c_\n
\,,\nonumber\\
\tilde{B}_{\m\n}&=&\p_\m \tilde{B}_\n -\p_\n \tilde{B}_\m\,,
\ea
where $\tilde{F}^a_{\m\n}=\sqrt{v} F^a_{\m\n}$ and $\tilde{B}_{\m\n}=\sqrt{v} B_{\m\n}$.

The electroweak Lagrangian is $\cL_{\rm ew} =\cL_{\rm YM} +\cL_\Phi -V(\Phi)$, with
\ba 
\cL_{\rm YM}&=&-\frac{1}{4}F^a_{\m\n}F_a^{\m\n} -\frac{1}{4}B_{\m\n}B^{\m\n}\,,\label{lym}\\
\cL_\Phi &=&-\left(\nabla_\m\Phi\right)^\dagger\left(\nabla^\m\Phi\right)\,,\label{hggk}\\
V(\Phi) &=& \frac{\l}{4}\left(\Phi^\dagger\Phi-\frac12 w^2\right)^2,\label{hggv}
\ea 
where $V(\Phi)$ is the Higgs potential providing a nonzero vacuum expectation value (VEV) to the Higgs doublet. 

To obtain a Standard Model whose free sector is stable in the integer picture, both $\lambda$ and $w$ must acquire a specific dependence on the measure weight $v(x)$. In particular, it is necessary that the VEV $w/{\sqrt{2}}$ depend on time and space via the relation
\be\label{V}
w(x)=\frac{w_0}{\sqrt{v(x)}}\,,
\ee
where $w_0$ is the (constant) value of the parameter in the usual theory. In fact, defining the Higgs scalar $\tilde\Phi:=\sqrt{v}\Phi$ in the integer picture, the minimum of the Higgs potential is at 
\be
\Phi^\dagger\Phi = \frac{w^2}{2}\quad\Rightarrow\quad \tilde\Phi^\dagger\tilde\Phi =\frac{w_0^2}{2}\,.
\ee
Despite having a varying minimum in the fractional picture and a constant one in the integer picture, spontaneous symmetry breaking leads to a constant Higgs mass in both pictures, provided we allow also $\tilde \lambda$ to vary. To show this, let the VEV be
\be 
\label{VEV}
\langle 0| \Phi |0\rangle = \frac{1}{\sqrt2} \left(\begin{matrix} 0 \\  w\\ \end{matrix}\right)
\quad\Rightarrow\quad \langle 0| \tilde\Phi |0\rangle = \frac{1}{\sqrt2} \left(\begin{matrix} 0 \\  {w_0}\\ \end{matrix}\right).
\ee
The mass terms ${\cal L}_M$ for the gauge fields are found by replacing $\Phi$ by its VEV in the kinetic term $\cL_\Phi$ for the Higgs isodoublet:
\be
\label{lmass1}
{\cal L}_M = -\frac18 w^2\left(\begin{matrix} 0 ,  1 \\ \end{matrix}\right)\!
\left(\begin{matrix}
g'{A}_\m^3 + g B_\mu   & g'({A}_\m^1-\rmi {A}_\m^2) \\
g'({A}_\m^1 +\rmi {A}_\m^2)   & -g'{A}_\m^3 + g B_\mu\\
\end{matrix}\right)^2\!\left(\begin{matrix} 0 \\  1 \\ \end{matrix}\right)\!.
\ee
To diagonalize the mass matrix, we introduce the picture-independent Weinberg angle $\theta_{\rm W}$:
\be
\label{thetaw}
\theta_{\rm W} := \tan^{-1}\frac{g}{g'} =\tan^{-1}\frac{g_0}{g'_0}\,,
\ee
and then denote, as usual,
\ba
W_\mu^\pm &:=&\frac{1}{\sqrt{2}}(A^1_\mu \pm \rmi A^2_\mu)\,,\label{W}\\
Z_\m &:=& \cos\theta_{\rm W} A^3_\m -\sin \theta_{\rm W} B_\m\,,\label{Z}\\
A_\mu &:=& \sin\theta_{\rm W} A^3_\m +\cos \theta_{\rm W} B_\m\,.\label{A}
\ea
The mass terms for the gauge fields read
\be
\label{lmass2}
{\cal L}_M = - M_W^2 W^{+\mu} W^-_{\mu} -\frac12 M_Z^2 Z^\mu Z_\mu\,,
\ee
where
\ba
\label{masses}
M_W &:=& \frac{g' w}{2} = \frac{g'_0 w_0}{2}\,,\\
M_Z &:=& \frac{g' w}{2 \cos \theta_{\rm W}} = \frac{g'_0 w_0}{2 \cos \theta_{\rm W}}\,.
\ea
Notice that, just like the Weinberg angle, also the boson masses do not depend on the picture 
(fractional or integer).

With the above settings, the electromagnetic coupling $e$ can be extracted by looking at the 
interaction between $A_\m$ and $A^a_\m$ in the Yang--Mills term:
\be
\label{elec}
e= g'\sin\theta_{\rm W} = g\cos\theta_{\rm W}\,.
\ee
Thanks to Eq.\ \Eq{g}, these relations are compatible with \Eq{e}.

Finally, to get the Higgs mass, we parametrize the Higgs doublet in the unitary gauge as
\be
\label{sigma1}
\Phi =\frac{1}{\sqrt{2}}\left(\begin{matrix} 0 \\  w +\sigma\\ \end{matrix}\right)\,,
\ee
so that the Higgs potential reads
\be
\label{hpot}
V(\Phi)=U(\s)=\sigma^2 \, \frac{\l w^2}{4} \left(1+\frac{\sigma}{2 w}\right)^2\,.
\ee
The global scale $\l(x)$ of the Higgs potential can be chosen in such a  way that the Higgs field $\s$
has the same constant mass in the fractional and integer picture. This is obtained by fixing
\be
\l(x)=  v(x)\,\l_0\,,
\ee
so that the Higgs mass reads
\be \label{hmass}
m_\s^2 =\frac{\l w^2}{2}= \frac{\l_0 w_0^2}{2}\,.
\ee
Overall,
\ba
U(\s) &=&\frac12m_\s^2\sigma^2+ \frac{\l w}{4} \s^3+  \frac{\l}{16}\s^4\\
      &=&\frac12m_\s^2\sigma^2+ \sqrt{v(x)}\frac{\l_0 w_0}{4} \s^3+ v(x)\frac{\l_0}{16}\s^4.\label{fundh}
\ea


\subsection{Multiscale electroweak model: Leptonic sector}\label{sec:lep}

After dealing with the gauge bosons and the Higgs field of the model, we turn our attention to leptons, considering for the sake of brevity only the electron and the electron neutrino. The left-handed fermions are placed in the weak isospin doublet $L=\left(\begin{matrix} \n_e\\ e_{\rm L}\\ \end{matrix}\right)$, whereas the right-handed electron is an isospin singlet $e_{\rm R}$. The hypercharge assignments are $Y_{\rm L}=1/2$ and $Y_{\rm R}=1$. Then, gauge covariant derivatives are
\ba
\nabla_\m L &=& \left(\cD_\m + \frac{\rmi}{2}g'\sigma_a {A}^a_\m +\frac{\rmi}{2}g  B_\m\right)L\,,\label{covL}\\
\nabla_\m e_{\rm R} &=& (\cD_\m +\rmi g  B_\m) e_{\rm R}\,.\label{covL2}
\ea
With this, we arrive at the free fermion Lagrangian in the fractional picture:
\be \label{lfree}
\cL_{\rm f} =\rmi\ove{e_{\rm R}}\g^\m\nabla_\m e_{\rm R} +\rmi\ove{L}\g^\m\nabla_\m L\,.
\ee

Combining Eqs.\ \Eq{W}--\Eq{A} and \Eq{covL}--\Eq{lfree}, we obtain
\ba \label{lf}
\cL_{\rm f} &=& \rmi \ove{{e}_{\rm L}}\g^\m\cD_\m{e}_{\rm L}+\rmi \ove{{e}_{\rm R}}\g^\m\cD_\m{e}_{\rm R} +\rmi\ove{\n}_e\g^\m\cD_\m\n_e \nonumber\\
 & & - g\ove{e}_{\rm R}\g^\m \cos\,\theta_{\rm W}\,A_\m e_{\rm R}\nonumber\\
 & & -\frac{1}{2}\ove{e}_{\rm L}\g^\m\left(g'\sin\,\theta_{\rm W} +g\cos\,\theta_{\rm W}\right)A_\m e_{\rm L}\nonumber\\
 & & +\frac{1}{2}\ove{\n}_e\g^\m\left(g'\sin\,\theta_{\rm W}-g\cos\,\theta_{\rm W}\right)A_\m\n_e\nonumber\\
 & & +\frac{1}{2}\ove{e}_{\rm L}\g^\m\left(g\sin\,\theta_{\rm W}-g'\cos\,\theta_{\rm W}\right)Z_\m e_{\rm L}\nonumber\\
 & & +\frac{1}{2}\ove{\n}_e\g^\m\left(g\sin\,\theta_{\rm W}+g'\cos\,\theta_{\rm W}\right)Z_\m\n_e\nonumber\\
 & & -\frac{1}{\sqrt{2}}g'\left(\ove{e}_{\rm L}\g^\m W^-_\m\n_e +\text{H.c.}\right)\nonumber\\
&=& \rmi \ove{{e}_{\rm L}}\g^\m\cD_\m{e}_{\rm L}+\rmi \ove{{e}_{\rm R}}\g^\m\cD_\m{e}_{\rm R} +\rmi\ove{\n}_e\g^\m\cD_\m\n_e \nonumber\\
 & & -e\,\ove{e}_{\rm R}\g^\m A_\m e_{\rm R}-e\,\ove{e}_{\rm L}\g^\m A_\m e_{\rm L}\nonumber\\
 & & +\frac{1}{2}\ove{e}_{\rm L}\g^\m\left(g\sin\,\theta_{\rm W}-g'\cos\,\theta_{\rm W}\right)Z_\m e_{\rm L}\nonumber\\
 & & +\frac{1}{2}\ove{\n}_e\g^\m\left(g\sin\,\theta_{\rm W}+g'\cos\,\theta_{\rm W}\right)Z_\m\n_e\nonumber\\
 & & -\frac{1}{\sqrt{2}}g'\left(\ove{e}_{\rm L}\g^\m W^-_\m\n_e +\text{H.c.}\right)\,,
\ea
where H.c.\ stands for Hermitian conjugate. In the second equality we used \Eq{elec}, which provides the correct electromagnetic coupling for both the left and right components of the electron, leaving the neutrino neutral.

Finally, we consider a Yukawa interaction, through which fermions acquire mass once the Higgs boson acquires an expectation value. In the fractional picture, it is defined as 
\ba 
\cL_{L\Phi} &=&-G_e\left(\ove{\n}_e,\ove{{e}}\right)_{\rm L}\Phi \,{e}_{\rm R}+\text{H.c.}\label{yuka}\\
&\to&-G_e\frac{w}{\sqrt{2}}\ove{{e_{\rm L}}}{e_{\rm R}}+\text{H.c.}\,,\label{emas}
\ea
where we have taken the VEV \Eq{VEV} and $G_e$ is the Higgs-lepton coupling. If we require the lowest-order electron mass $m_e$ to be constant in the integer picture at the tree level, then
\be \label{ge}
G_e(x)=\sqrt{v(x)}\,G_{0e}
\ee
and $m_e =w G_e=w_0 G_{0e}$ in Eq.\ \Eq{emas}.


\subsection{Multiscale chromodynamics: Inclusion of quarks}\label{sec:quarks}

The inclusion of quarks is straightforward. Without loss of generality, we shall only consider the first quark family,  $(u,d)$, that belongs to the fundamental representation of the
$SU(3)$ (color) gauge group, with generators given by the $3\times 3$ Gell-Mann matrices $\l^a$, $a=1,\ldots,8.$\footnote{To avoid a proliferation of symbols, we shall adopt Latin indices $a, b, c, \ldots$ to enumerate gauge generators. It will be clear from the context whether the index refers to an $SU(3)$ or $SU(2)$ generator.}  Color gauge potentials will be denoted by $C^a_\m$ and the strong coupling by $g_s$. The relation between the strong coupling in the fractional picture and the usual coupling constant $g_{0s}$ in the integer picture is $g_s= \sqrt{v}\, g_{0s}$. The first quark family $(u,d)$ forms a left-handed Weyl spinor ${\rm q}_i$, $i=1,2=u,d$ under $SU(2)$ gauge transformations. In the same way, we shall introduce the antiquarks $\bar u$ and $\bar d$ which are singlets under $SU(2)$. The bar over $\bar u$ and $\bar d$ are part of the definition of the field and it does not imply any sort of conjugation. Hypercharge assignments for the quarks are $1/6$ for the Weyl doublet q and $-2/3$ and $1/3$ for the singlets $\bar u$ and $\bar d$, respectively. Consequently, the covariant derivatives read
\ba
(\nabla_\m {\rm q})_{\a i} &=& \cD_\m  {\rm q}_{\a i} + \rmi g_s C^a_\m (\l^a)_\a^{\ \b} {\rm q}_{\b i}+ \frac{\rmi}{2}g'{A}^a_\m (\sigma_a)_{i}^{\  j} {\rm q}_{\a j}\nonumber\\
&&  +\frac{\rmi}{6}g  B_\m {\rm q}_{\a i}\,,\label{covQ}\\
(\nabla_\m \bar u)^\a &=&\cD_\m  \bar u^\a + \rmi g_s C^a_\m (\l^a)^\a_{\ \b} \bar u^\b -\frac{2 \rmi}{3}g  B_\m \bar u^\a\,,\label{covbarU}\\
(\nabla_\m \bar d)^\a &=&\cD_\m  \bar d^\a + \rmi g_s C^a_\m (\l^a)^\a_{\ \b} \bar d^\b +\frac{ \rmi}{3}g  B_\m \bar d^\a\,.\label{covbarD}
\ea

Then, the kinetic term for quarks in the fractional picture reads
\be \label{lquarks}
\cL_{\rm q} =\rmi {\rm q}^{\dagger \a i} \bar \s^\m ( \nabla_\m {\rm q})_{\a i} +\rmi \bar u^{\dagger}_\a \bar \s^\m ( \nabla_\m \bar u)^\a +\rmi \bar d^{\dagger}_\a \bar \s^\m ( \nabla_\m \bar d)^\a \ ,
\ee
where $\bar \s^\m = (\mathbbm{1}, - \s^a)$ and $\mathbbm{1}$ is the $2\times 2$ identity matrix. A mass term for quarks cannot be included because there is no gauge-group singlet contained in any of the products of their representations, as is well known. Consequently, mass terms for quarks arise only after spontaneous symmetry breaking. To this purpose, we introduce the Yukawa couplings between quarks and Higgs field,
\be \label{LYuk}
{\cal L}_{\rm Yuk}= y'\epsilon^{ij}\Phi_i {\rm q}_{\a j}\bar u^\a-y''\Phi^{\dagger i} {\rm q}_{\a i}\bar d^\alpha +{\rm H.c.}\,,
\ee
where $\Phi^i$ are the two components of the Higgs field $\Phi$ and $y',\, y''$ are two couplings in the fractional picture, related to the constant couplings of the integer picture by $\sqrt{v}\,y'_0,\, \sqrt{v}\,y''_0$.

In the unitary gauge, the Higgs field has the form \Eq{sigma1} and \Eq{LYuk} reads
\be \label{LYuk2}
{\cal L}_{\rm Yuk}= - \frac{\s+w}{\sqrt{2}}\bigl[ y' ( u_\a \bar u^\a+ \bar u^\dagger_\a u^{\dagger \a} )+ y'' ( d_\a \bar d^\a+ \bar d^\dagger_\a d^{\dagger \a} )\bigr].
\ee
Defining a pair of Dirac Fermions $\Psi_u$ and $\Psi_d$ for the up and down quarks
as
\be
\label{ud}
\Psi_u = \left(\begin{matrix} u_\a \\ \bar u^\dagger_\a\\ \end{matrix}\right), \quad \Psi_d = \left(\begin{matrix} d_\a \\ \bar d^\dagger_\a\\ \end{matrix}\right),
\ee
we immediately recognize in \Eq{LYuk2} the Dirac mass terms for the $(u,d)$ quarks,
\be
\label{udmasses}
m_u = \frac{y' w}{2} = \frac{y'_0 w_0}{2}\ , \qquad m_d=\frac{y'' w}{2}=\frac{y''_0 w_0}{2}\,.
\ee
As expected, the masses in the fractional picture are equal to the masses in the integer picture. The remaining $\s$-dependent terms in \Eq{LYuk2} provide the Yukawa couplings between the Higgs and up and down quarks.

 
\subsection{Classical symmetries}\label{simw}

It has been recognized since early works that multiscale field theories break Poincaré invariance due to the explicit breaking of translation and rotation invariance by the measure. An inspection of the Poincaré algebra refines this conclusion. 

The first step is to define the generators of the ``fractional'' generalization of translations, rotations, and boosts in the theory with weighted derivatives, the only difference with respect to the usual case being in the type of derivatives involved. Then, one can derive the transformation rules for tensor fields. This was done in \cite{frc6} for a scalar field $\phi$, where it was found that $\phi$ is a scalar density of weight $-1/2$ with respect to the measure. This is obvious from the fact that $\tilde\phi=\phi/\sqrt{v}$ is an ordinary scalar when the theory is recast in the integer picture. Actually, the results of \cite{frc6} apply to fields of any tensorial rank: in the construction of the preceding sections, the same density weight $\sqrt{v}$ connects rank-0 and rank-1 fractional fields with ordinary fields. The same rule applies to rank-2 and higher tensors \cite{frc11}.

The second step is to check what type of algebra the fractional Poincaré generators (which, we stress again, do not represent ordinary transformations) obey. It turns out that, for the scalar theory with weighted derivatives, it is the standard Poincaré algebra only in the absence of nonlinear interactions \cite{frc6}. When nonlinear interactions are switched on, the algebra is deformed; for instance, the Poisson bracket between the fractional momentum and the Hamiltonian does not vanish. It is easy to extend this conclusion to the full Standard Model defined here: the Poincaré algebra holds in all sectors except the Higgs, which is the only one containing nonlinear interactions (see Sec.\ \ref{sec:fvi} for related comments). As we will discuss in the next section, the physical symmetries of the system are those valid in the fractional picture. Therefore, even if the action is formally the standard one in the integer picture, one can conclude that standard Poincaré invariance is not a classical physical symmetry of the theory.

Having discussed the transformation properties of the fields and violation of local Poincaré symmetries, let us consider discrete Lorentz transformations (charge conjugation C, parity P, and time reversal T). Recall that the requirement of having a nonnegative-definite measure implies that the geometric coordinates are odd under reflection $q^\mu(-x^\mu)=-q^\mu(x^\mu)$. Since the measure weight \Eq{genmea} is even in the coordinates, the classical theory is invariant under parity and time-reversal transformations PT. The presence of measure weights does not affect the charge properties of spinors, so also C is preserved.

Quantum symmetries will be commented upon in Sec.~\ref{sec:conc}.


\section{Physics of the theory with weighted derivatives}\label{sec:vs}

After building the Standard Model in the theory with weighted derivatives, we turn to analyze its physical consequences. In particular, there are some pending questions left from previous studies and mentioned in Sec.\ \ref{intro}. Is the quantum theory well defined? Can accelerators unravel an anomaly in the dimension of spacetime? 

We have already formulated the problem of choice between the fractional and the integer picture, which is relevant for the issue of the observational consequences of the theory. In Sec.\ \ref{sec:qft}, we will recall (also with some new results compared to \cite{frc9}) why it can be difficult or even impossible to formulate a predictive perturbative quantum field theory in spacetimes with weighted derivatives. Next, in Sec.\ \ref{sec:fvi} we will see how problems disappear in the case of the Standard Model. Combining the results of this paper with those of \cite{frc11,frc14}, in Sec.\ \ref{versus} we eventually show that the theory with weighted derivatives is self-consistent, well defined as a quantum field theory and nontrivial. However, signatures of an anomalous dimension must be looked for either in the electromagnetic sector or away from accelerators, in experiments of atomic physics or in the realm of astrophysics and cosmology.

 
\subsection{Quantum interactions}\label{sec:qft}

In principle, the triviality of the integer picture can be easily broken in the presence of nonlinear interactions: couplings $\tilde\l_i(x)$ acquire a nontrivial measure dependence which is impossible to absorb. However, nonconstant couplings make the quantum field theory hard to deal with. In the multiscale scenario with weighted derivatives, the nonconservation of the energy-momentum tensor in the fractional picture implies that momentum along the $\mu$ direction spreads out for a generic $\alpha_\mu$. For very special values of $\alpha_\mu$, momentum is perturbatively conserved at the quantum level at least if the perturbative series is truncated at any finite order \cite{frc9} but, unfortunately, a proof of conservation at all orders is missing due to the difficulty in formulating Feynman rules even in this special case. Let us now recall the problem in the example of a scalar-field theory in multiscale Minkowski spacetime in $D$ topological dimensions, with Lagrangian $\cL=-\cD_\mu\phi\cD^\mu\phi/2-V(\phi)$ and potential
\be\label{monpot}
V(\phi)=\frac12 m^2\phi^2+\l_0\frac{\phi^n}{n!}\,,\qquad n=3,4,\dots\,.
\ee
For this theory, the matrix element $\langle {\rm f}\vert {\rm i} \rangle$ on two-particle states has been computed in \cite{frc9} for $n=3$ and the fractional measure weight \Eq{smea}. Here we extend these results to arbitrary $n$ and the more realistic measure profile \Eq{binomialm}, which represents a geometry with a $D$-dimensional infrared limit.

In the integer picture, the model can be recast as an ordinary field theory but with potential $\tilde V(\tilde\phi)=m^2\tilde\phi^2/2+\l(x)\tilde\phi^n/n!$, where $\tilde\phi=\sqrt{v}\phi$ and $\l(x)=\l_0[v(x)]^{1-n/2}$ \cite{frc6}. The tree-level $n$-valent vertex in the integer picture is then 
\ba
\tilde{\cal V}(k_1,\ldots , k_n) &=& \rmi \int\rmd^D x \,\l(x)\,\rme^{\rmi x\cdot k_{\rm tot}}\nonumber\\
																 &=& \rmi\l_0\prod_\mu\int\rmd x^\mu\,[v_\mu(x^\mu)]^{1-n/2}\rme^{\rmi x_\mu k_{\rm tot}^\mu}\,,\nonumber\\
\ea
where $k_{\rm tot}^\mu=\sum_{i=1}^n k^\mu_i$ and Einstein sum conventions are \emph{not} used in the second line. For the measure weight \Eq{binomialm}, we can expand the integral in an infrared and late-time regime $v_\mu\simeq1+\delta v_\mu$, where $\delta v_\mu\propto |x^\mu/\ell_*^\mu|^{\a_\mu-1}\ll 1$:
\ba
\tilde{\cal V}(k_1,\ldots , k_n)   &=& \tilde{\cal V}_0+\delta\tilde{\cal V}+O(\delta v^2)\,,\label{V0}\\
\tilde{\cal V}_0(k_1,\ldots , k_n) &=& \rmi\l_0\prod_\mu\int\rmd x^\mu\,\rme^{\rmi x_\mu k_{\rm tot}^\mu},\\
\delta\tilde{\cal V}(k_1,\ldots , k_n) &=& -\left(\frac{n}{2}-1\right)\rmi\l_0\sum_\mu\delta\tilde{\cal V}_\mu\,,\\
\delta\tilde{\cal V}_\mu(k_1,\ldots , k_n) &=&\int\rmd x^\mu\,\left|\frac{x^\mu}{\ell_*^\mu}\right|^{\a_\mu-1}\rme^{\rmi x_\mu k_{\rm tot}^\mu}.\label{deV}
\ea
While in our case $v^{1-n/2}\simeq 1-(n/2-1)\delta v$, in \cite{frc9} we only have $(\delta v)^{1-n/2}$. This leads to an important difference between \Eq{deV} and the vertex in \cite{frc9}, apart from the value and sign of the prefactor: the exponent in the integrand. Here, for each direction (label $\mu$ omitted) we have $\a-1$, while in \cite{frc9} one has $\b-1:=(\a-1)(1-n/2)$. Consequently, the allowed values of $\a_\mu$ for which one can obtain user-friendly Feynman rules will differ with respect to \cite{frc9} [see Eq.\ (20) therein].

Equation \Eq{V0} is the standard vertex $\tilde{\cal V}_0=\rmi \l_0 (2\pi)^D\delta(k_{\rm tot})$, where $\delta$ is the $D$-dimensional Dirac delta. Equation \Eq{deV} conserves momentum only for special values of the exponent. If $\a_\mu=2l_\mu+1$ with $l_\mu\in\mathbbm{N}$, then
\be\label{devkt}
\delta\tilde{\cal V}_\mu = \frac{2\pi}{(\ell_*^\mu)^{2l_\mu}}\frac{(-1)^{l_\mu}}{(2l^\mu)!}\,\delta^{(2l_\mu)}(k_{\rm tot}^\mu)\,,
\ee
where $\delta^{(2l_\mu)}$ is the derivative of order $2l_\mu$ of the one-dimensional delta.

At this point, one recognizes three major problems with \Eq{devkt}. First and foremost, due to the presence of derivatives acting on the delta distribution, it is not guaranteed that the effective vertex from the infinity of loop diagrams will have support at $k_{\rm tot}=0$. Second, it is difficult to compute loop diagrams with vertices \Eq{devkt} unless one further assumes that only one direction $\bar\mu$ is anomalous, while $v_\mu= 1$ for all the other $\mu\neq\bar\mu$. This assumption seems a necessary technical demand but it reduces the generality of the model drastically. Third, even ignoring the previous two issues it is hard to embed the model in multiscale spacetimes, where the fractional exponents $\a_\mu$ take values in the interval $(0,1]$ \cite{frc1,frc2}.\footnote{In the ultraviolet, for $\a_\mu>1$ one would obtain a spacetime dimension larger than in the infrared, a possibility not unphysical but not usually realized in quantum gravity, either. For $\a_\mu<0$, the dimension in the ultraviolet may be ill defined (negative definite). If only some of the exponents take large or negative values, then one can still obtain a well-defined spacetime dimensionality across all scales and the interval $(0,1]$ can be slightly extended. This may not be true in other multiscale theories \cite{frc1}.} The nontrivial values $\alpha_\mu=3,5,7,\dots$ allowed here do not fit in such a range.

 
\subsection{Standard Model in the integer picture}\label{sec:fvi}

We now apply the above considerations to the Standard Model built in Sec.\ \ref{ewwd}. The first observation we make is that, contrary to expectations of Sec.\ \ref{sec:qft}, the integer picture is trivial, i.e., the model is an ordinary quantum field theory.

Thanks to the field redefinition \Eq{ca}, we have been able to recast the system \Eq{lagyan} with weighted derivatives and spacetime-dependent couplings as the system \Eq{tilL} where the Lagrangian $\tilde{\cal L}:=v {\cal L}$ has ordinary derivatives and constant couplings. The constancy of all the couplings is a consequence of Eq.\ \Eq{g}. Since the system in the fractional picture \Eq{lagyan} is the same as the one in the integer picture \Eq{tilL}, there is no nontrivial information in \Eq{lagyan}. A similar exact equivalence between the fractional and the integer picture was shown in \cite{frc8} for electrodynamics. Both electrodynamics and its weak-strong-force extension are interacting theories, as fermions couple with gauge fields via three-legged vertices of the form $\bar \psi (g\cA) \psi$. Homogeneity of the covariant derivative (i.e., the derivative and gauge terms must scale in the same way) and the requirement of a clear notion of gauge invariance forced us to assume a coupling $g(x)$ with a specific spacetime dependence. However, the profile $g(x)$ is such that interactions of the type ``$B^2 A$'' are trivial because they are at most quadratic in the fields $A$ and $B$. The two spinor fields in $\bar \psi (g\cA) \psi$ reabsorb the weight $v$ in the integration measure, while the bosonic vector combines with the coupling so that $g\cA = g_0\tilde\cA$. Thus, the vertex in the integer picture has no $v(x)$ factors.

The Higgs sector is also trivialized in the integer picture. After the field redefinition $\tilde\s=\sqrt{v}\s$, the potential \Eq{fundh} becomes the usual one:
\be 
v\,U(\s)=\tilde U(\tilde\s)=\frac12m_\s^2\tilde\s^2+ \frac{\l_0w_0}{4} \tilde\s^3+ \frac{\l_0}{16}\tilde\s^4.\label{tfundh}
\ee
Order by order, the measure dependence of the couplings is exactly reabsorbed. This phenomenon is not possible with a potential with only one nonlinear term, as in \Eq{monpot}; the problems found in \cite{frc9} are thus avoided. 

Notice that the cancellations leading to \Eq{tfundh} are not an accident due to the mutual dependence of the couplings. Their main cause is the requirement that shifts of the field $\s$ be homogeneous in the anomalous scaling, i.e.,
\be\label{shif}
\s(x)=\s'(x)-\s_0(x)
\ee
in the fractional picture implies $\tilde\s(x)=\tilde\s'(x)-\tilde\s_0$ in the integer picture, where $\tilde\s_0=\sqrt{v(x)}\s_0(x)$ [this is the equivalent of \Eq{V}].\footnote{In turn, homogeneity in the shift implies constancy of the mass.} To see this, consider the potential
\be\label{Wph}
U(\s) = a_0+a_1\s+a_2\s^2+a_3\s^3+a_4\s^4
\ee
instead of \Eq{fundh}. The constant and linear terms can be eliminated by the shift \Eq{shif}. Substituting \Eq{shif} into \Eq{Wph}, the coefficients of the constant and linear terms vanish if, and only if, $a_0'=a_0-a_1\s_0+a_2\s_0^2-a_3\s_0^3+a_4\s_0^4=0$ and $a_1'=a_1-2a_2\s_0+3a_3\s_0^2-4a_4\s_0^3=0$. Plugging these relations back into the potential and taking into account the overall measure prefactor $v$, we have
\be
a_n\propto \frac{1}{v(x)\,\s_0^n(x)}\propto [v(x)]^{\frac{n}{2}-1}, 
\ee
precisely as in \Eq{fundh}.

 
\subsection{Fractional versus integer picture}\label{versus}

Since all couplings are constant in the integer picture, the quantum field theory is well defined and manageable at all perturbative orders. However, the inevitable conclusion is that the theory with weighted derivatives is not multiscale at all in the integer picture: it is formally equivalent to the ordinary Standard Model in Minkowski spacetime. Also, it is easy to convince oneself that any measurement of time or space intervals will be the same in both pictures: when moving back to the fractional picture, one undoes the field redefinition \Eq{vpph} but coordinates remain untouched. Therefore, even if intervals are calculated theoretically with a nontrivial measure, the measurement units remain the same. 

On top of all this, we have seen that the couplings in the weak sector combine in a neat way eliminating the measure dependence. In Sec.\ \ref{sec:feynman} we will describe the example of the lifetime of the muon and check that the usual prediction is obtained, even in the fractional picture. The strong sector follows a similar fate. Therefore, a \emph{flat} multiscale world with weighted derivatives \emph{completely and solely} described by weak and strong interactions cannot be tested in particle accelerators.

Does this mean that the theory is trivial? The answer is No.  As said in Sec.\ \ref{sec:pic}, a trivial integer picture does not necessarily imply that there is no observable consequence of having a multiscale geometry. The two caveats ``flat'' and ``completely and solely'' forbid to draw a similar conclusion for all multiscale systems with weighted derivatives. Couplings which are measured directly can bear the marks of a multiscale geometry. The electric charge $Q(t)$ in \Eq{ech} is one such case \cite{frc8} and the Lamb-shift example of Sec.\ \ref{lashi} will reiterate the point. The caveat on flatness of the background covers many subtle points, which will be described in the following.

\subsubsection{Without gravity}

Systems described by statistical or particle mechanics can feel the distinct presence of an anomalous scaling, via quantities such as the density of states per unit energy. Examples are the random motion of a molecule \cite{frc7}, the dynamics of a relativistic particle \cite{frc10} and the black-body radiation spectrum \cite{frc14}, all processes with a characteristic energy much smaller than that in the center of mass of subatomic scattering events. This does not mean, of course, that multiscale effects are more prominent at low energies: the corrections to standard results are progressively smaller as the energy decreases, and effects of the anomalous geometry are virtually undetectable at mesoscopic scales. Rather, the reason why statistical and particle-mechanics systems yield nontrivial predictions is that they are not subject to requirements as severe as those we imposed on a quantum field theory, namely the constancy of masses (to allow for a manageable quantum perturbative treatment) and the enforcement of gauge symmetries. Such constraints, purely dictated by the way we are able to deal with quantum fields, limit the way the field-theory degrees of freedom couple nonlinearly. On the other hand, statistical and particle-mechanics settings are intrinsically nonlinear, either through the stochastic interaction of a degree of freedom with the environment (as in the multiscale Brownian motion of a particle \cite{frc7}), or by definition of the action (as for the relativistic particle \cite{frc10}), or via the collective description of microscopic degrees of freedom (as in the frequency distribution of a thermal bath of photons \cite{frc14}).\footnote{We have to mention that all these systems have another property in common: they treat space coordinates $x^i$ and time $t$ (or diffusion time $\s$ \cite{frc7}, or proper time $\tau$ along a geodesic \cite{frc10}) on a different footing. This gives rise, in general, to an ambiguity in the measure weight along different directions, which the theory can constrain only by combining the information of all these different systems. For example, by themselves stochastic methods are unable to fix the anomalous weight $v_0(\s)$ along diffusion time and there are different possible values for the spectral dimension $\ds$ of spacetime \cite{frc7}. In parallel, in nonrelativistic mechanics \cite{frc5} the weight $v_0(t)$ along the particle trajectory can be arbitrarily chosen among the allowed shapes dictated by fractal geometry (Sec.\ \ref{sec:mea}). When constraining the weights $w_\mu(\tau)$ for a relativistic particle by matching the action with the nonrelativistic limit, one is forced to conclude that $v_0(\s)=v_0(t)=1=w_\mu(\tau)$ and that the time direction is ordinary \cite{frc10}. This fixes the ambiguity in stochastic processes and determines the spectral dimension to be the ordinary one, $\ds=D$. In turn, a nonanomalous time would exclude the variation of the electric charge and of the fine-structure constant found in \cite{frc8}, Eq.\ \Eq{oldco}. In \cite{frc10}, it was suggested to pick the relativistic action as the basic definition for the dynamics of a single particle, and to simply accept its nonrelativistic limit as it is. Then, one does not have to match such limit with the less fundamental construction of \cite{frc5}, the weights are unconstrained and so is the spectral dimension and the geometry of time. Then, the dynamics of charged particles does not admit a trivial integer picture. This is not in contradiction with the results on the Standard Model obtained in \cite{frc8} and in the present paper. Even regarding quantum field theory as the fundamental framework where all the rest of the physics should ideally stem (via thermodynamical or nonrelativistic approximations), a standard Standard Model in the integer picture does not imply a standard particle mechanics or a standard spectral dimension in the integer picture. The above-mentioned nonlinearities intervening in the limit from field theory to particle and stochastic mechanics make such transition highly nontrivial.

The presence of ambiguities in the formulation of certain atomic-mesoscopic sectors may suggest that the theory with weighted derivatives is not a fundamental description of Nature but, rather, an effective model valid in regimes where effects of the putative fundamental anomalous geometry become apparent. This possibility depends on the overall control we can exercise on the theory and it is not excluded by our present level of knowledge.}


\subsubsection{With gravity}\label{withg}

Acting as the Devil's advocate, one might object that a choice of frame is a somewhat weak expedient to save the Standard Model with weighted derivatives from a death sentence. This point of view would disregard the fundamental change of perspective entailed in multiscale geometries, where multiscale measurements are performed with multiscale instruments by multiscale observers. But even granting that, by definition, there is no physical equivalence of the fractional and the integer picture, the almost triviality of the Standard Model in the theory with weighted derivatives is somewhat disappointing. After all, one would have liked to constrain a new spacetime geometry in all possible sectors of physics, especially in one which is subject to the most severe precision tests in science and is undergoing the recent and exciting developments of LHC. However, until now we have ignored a fundamental factor of discrimination between the fractional and the integer picture. This factor is model independent, it neutralizes the ``equivalence of frames'' staunch viewpoint and it becomes active when gravity joins the game and matter is coupled to a generic nonflat background with metric $g_{\mu\nu}$.

Consider the analogy of a similar problem of choice between the Einstein and the Jordan frame in scalar-tensor theories. The two frames are physically equivalent both classically and at the quantum level to first order in perturbation theory (also in a cosmological sense), but they differ in a nonlinear quantum regime. At that point, a choice of frame is necessary according to some criterion. For instance, one might regard the Jordan frame as the fundamental one because it is the frame where matter follows the geodesics. Another example of frame choice solved by a careful definition of the theory is the class of varying-speed-of-light models (see again \cite{frc8} for a discussion and a comparison with multiscale spacetimes).

Similarly, in the integer picture the multiscale theory with weighted derivatives is not general relativity with minimally coupled matter, and one can never trivialize the theory to the ordinary one as in the flat case \Eq{fip}. The gravitational dynamics of the theory with weighted derivatives was studied in \cite{frc11}. There are two versions of the gravitational sector. One has a standard gravitational field and the action is the same as the multiscale theory with ordinary derivatives:
\ba
S_g[v,\phi^i] &=&\frac{1}{2\kappa^2}\int\rmd^Dx\,v\,\sqrt{-g}\,\left[R-\om(v)\p_\mu v\p^\mu v-U(v)\right]\nonumber\\
&&+S[v,\phi^i]\,,\label{Sg2}
\ea
where $\om$ and $U$ are functions of the weight $v$, $R$ is the ordinary Ricci scalar and $S[v,\phi^i]$ is the matter contribution. Even setting $\om=0=U$, the gravitational sector is not the Einstein--Hilbert action, due to the presence of $v$. Absorbing weight factors into the matter fields $\phi^i$ with the picture change \Eq{vpph} requires a redefinition of the metric $g_{\mu\nu}\to\tilde{g}_{\mu\nu}$. Indeed, one can go to the integer picture (Einstein frame) where the gravitational action is $\propto \int\rmd^Dx\,\sqrt{-\tilde g}\,\tilde R$ but not without reintroducing nontrivial terms $\tilde\om\neq 0\neq \tilde U$ for the measure weight. These terms affect the cosmic evolution \cite{frc11}. The equations of motion are different from those in an ordinary scalar-tensor theory, since $v$ is not a scalar field and the action is not varied with respect to it.

The other version of the gravitational sector is more interesting, since the metric is not covariantly conserved [$\nabla_\s g_{\mu\nu}=\mp (\p_\s\ln v^\beta) g_{\mu\nu}$, where $\beta$ is a constant] and the geometry corresponds to a Weyl-integrable spacetime. The total action reads
\ba
S_g[v,\phi^i] &=&\frac{1}{2\kappa^2}\int\rmd^Dx\,v\,\sqrt{-g}\left[{\cal R}-\om\cD_\mu v\cD^\mu v-U(v)\right]\nonumber\\
&&+S[v,\phi^i]\,,\label{eha}
\ea
where ${\cal R}$ is the Ricci scalar constructed with weighted derivatives of order 0 (ordinary derivatives) and $\b$ [$\b=1/2$ in \Eq{wder} and $\b=1$ in \Eq{chD}] \cite{frc11}. As in the model \Eq{Sg2}, a change of picture does not lead to standard general relativity plus matter and the dynamics is different from (and much more constrained than) that of scalar-tensor scenarios in both frames.

Again, we should be careful about the issue of the physical (in)equivalence between the fractional and the integer picture. As for scalar-tensor models, from a simple visual inspection of the actions one cannot conclude that the Jordan and Einstein frames define different physics. What matters are the physical observables. For scalar-tensor theories in a classical cosmological homogeneous setting, the two frames are equivalent \cite{DeS10,ChY}, while a similar result does not hold for the multiscale theory with weighted derivatives since the fractional picture is postulated to be the fundamental frame. At any rate, the homogeneous classical cosmology of the multiscale theory is physically distinguishable from the usual one even in the integer picture (Einstein frame), since $\tilde\om\neq 0\neq \tilde U$. Moreover, at the quantum inhomogeneous level the physical equivalence between the Jordan and Einstein frames in scalar-tensor theories is broken \cite{Cho97,HuNi,ArPe,NiPi}. The same is true for the multiscale case.

To summarize, the multiscale field theory with weighted derivatives is self-consistent, predictive in all its sectors (particle phenomenology, cosmology, and so on) and can be physically told apart from its ordinary counterpart by appropriate measurements taking place in the electromagnetic sector and, more generally, at atomic or higher scales. Let us see an example in quantum electrodynamics: the well-known Lamb shift effect.


\subsection{Lamb shift}\label{lashi}

Since the only sector in the Standard Model where the theory with weighted derivatives produces observable exotic consequences is electrodynamics, we pick one of the most precise and accurate QED experiments available: the measurement of the Lamb shift. This will be more than enough to test the theory in the context of the Standard Model of quantum particles. In this section, we specialize to $D=4$ topological dimensions. 

According to Bohr's model, the spectrum of the electron in the hydrogen atom depends only on the principal quantum number $n$. Quantum field theory corrects this result. The emission and absorption of virtual photons by electrons and the production of virtual electrons in internal photon lines in Feynman diagrams give rise to a splitting of the spectral lines of different spin orbitals $l$ and, in particular, a shift in the energy of the ${}^2P_{1/2}$ state ($n=2$, $l=1$) with respect to the ${}^2S_{1/2}$ state ($n=2$, $l=0$). The measurement of this shift is one of the precision tests of quantum electrodynamics and has by now been verified for a number of light hydrogenic atoms (hydrogen, deuterium D, helium ion He${}^+$, muonium and positronium) \cite{EGS,Kar05}. For instance, the measured shift $\Delta E= E_{2S}-E_{2P}$ between the $2S$-$2P$ levels of hydrogen is \cite{Sch99}
\be
\Delta E      = 1057.8446(29)h\,\text{MHz}=4.37489(1)\times 10^{-6}\,\text{eV}\,,\label{hyd}
\ee
where $h$ is Planck's constant, we used the conversion $1\,\text{MHz}\times h\approx 4.13567\times 10^{-9}\,\text{eV}$ and the numbers in round brackets denote the first nonzero digits of the $1\s$-level experimental error and apply to the last figure(s) given in the number. A very close value has been found for the $2S$-$2P$ Lamb shift of deuterium, $\Delta E^{\text{D}} = 1059.234(3)h\,\text{MHz}$ \cite{Sch99}, while for ionic helium $\Delta E^{\text{He}\!^+} = 14041.1(2)h\,\text{MHz}$ \cite{Wij00}. The theoretical values predicted by quantum electrodynamics are all in excellent agreement with these observations.

The theoretical prediction for the Lamb shift consists of a sum of various contributions, including radiative corrections, form factors, two-particle recoil, and so on. 
Since we want to make an order-of-magnitude estimate of multiscale effects at scales larger than $t_*$ and $\ell_*$, it is sufficient and self-consistent to retain only leading-order terms in the fine-structure constant, which is the only source of such effects.

The leading contributions to the energy level $E_{n,l,j}$ are: (a) one-loop radiative insertions in the electron line and the Dirac form-factor contribution, (b) the contribution of the Pauli form factor $F_2$ and (c) the one-loop correction from the polarization operator. In ordinary quantum electrodynamics, one has \cite{EGS}
\be 
E_{n,l,j} = E_{\rm rad} + E_{F_2} + E_{\rm pola}\,,
\ee
with
\bs \label{lambc}\ba
E_{\rm rad} &=& \Bigg\lbrace \left[\frac{1}{3}\ln\frac{m}{m_r\,(Z\tilde\a_\textsc{qed})^{2}}+\frac{11}{72}\right]\delta_{l0}\nonumber\\
&& -\frac{1}{3}\ln k_0(n,l)\Bigg\rbrace \frac{4\tilde\a_\textsc{qed} m (Z\tilde\a_\textsc{qed})^4}{\pi\, n^3}\left(\frac{m_r}{m}\right)^3,\nonumber\\\\
E_{F_2} \Big\vert_{l=0} &=& \frac{\tilde\a_\textsc{qed}(Z\tilde\a_\textsc{qed})^4 m}{2\pi\,n^3}\left(\frac{m_r}{m}\right)^3,\\
E_{F_2}\Big\vert_{l\neq 0} &=& \frac{\tilde\a_\textsc{qed}(Z\tilde\a_\textsc{qed})^4m}{ 2\pi\,n^3}\nonumber\\
& & \times\frac{j(j+1)-l(l+1)-{3}/{4}}{l(l+1)(2l+1)}\left(\frac{m_r}{m}\right)^2,\\
E_{\rm pola} &=& -\frac{4\tilde\a_\textsc{qed}(Z\tilde\a_\textsc{qed})^4m}{15\pi\, n^3}\left(\frac{m_r}{m}\right)^3\delta_{l0}\,,
\ea\es
where $\ln k_0(n,l)$ is the Bethe logarithm (a computable function of the principal and orbital quantum numbers), $Z$ is the atomic number of an atom with nucleus mass $M$, $m$ is the electron mass, $m_r=mM/(M+m)$ is the reduced mass and $\tilde\a_\textsc{qed}=e_0^2/(\hbar c)$ is the fine-structure constant ($e_0$ being the electric charge), denoted like this in order to distinguish it from the fractional charge $\a$. For the $2S_{1/2}$-$2P_{1/2}$ Lamb shift of hydrogen, $Z=1$ and
\ba
\Delta E &=& E_{2,0,1/2}-E_{2,1,1/2}\nonumber\\
&=& \left[\ln\frac{m k_0(2,1)}{m_r\,\tilde\a_\textsc{qed}^2k_0(2,0)}+\frac{19}{30}+\frac{m}{8 m_r}\right]\frac{\tilde\a_\textsc{qed}^5 m}{6\pi}\left(\frac{m_r}{m}\right)^3.\nonumber\\\label{delam}
\ea

In the theory with weighted derivatives, these formul\ae\ are readily obtained in the integer picture. When converting them to the fractional picture, the energies on the left-hand sides and the masses on the right-hand sides remain unaffected but the fine-structure constant acquires a time dependence stemming from the observed electric charge \Eq{ech} on the measure in the time direction \cite{frc8}. In our units, in the fractional picture one has
\be
\a_\textsc{qed}(t)=Q^2(t)=\frac{e_0^2}{v(t)}=\frac{\tilde\a_\textsc{qed}}{v(t)}\,.
\ee
The fine-structure constant is time dependent and space independent, but not by virtue of an arbitrary assumption: it is a consequence of the conservation law of Noether currents in the theory; see \cite{frc8} for details. For the binomial profile \Eq{mt2}, we get
\be\label{plugal}
\tilde\a_\textsc{qed}=\a_\textsc{qed}(t)\,v_*(t)= \a_\textsc{qed}(t)\,\left(1+\left|\frac{t}{t_*}\right|^{\a_0-1}\right).
\ee
Here, $\a_\textsc{qed}(t)$ is the observed fine-structure constant, measured at some time $t$ which depends on the experiment. For measurements of the spectra of cosmologically distant objects such as quasars, $t$ is the cosmic time that passed since the emission of light of such objects. For a particle-physics experiment, $t$ is the characteristic time $t_\textsc{qed}\sim 10^{-21}\mhyp 10^{-16}\,{\rm s}$ of the electromagnetic interaction, corresponding to the lifetime of unstable particles that decay via such interaction. Because of \Eq{plugal}, at any given time $t=t_{\rm exp}$, $\a_\textsc{qed}(t_{\rm exp})<\tilde\a_\textsc{qed}$ and the effect of the multiscale geometry is a change in magnitude of the measured fine-structure constant, always smaller than the usual one.

Plugging \Eq{plugal} into \Eq{delam} and expanding for $t\gg t_*$ to first order, one obtains
\ba
\Delta E &\simeq& \Delta E^{(0)}+\Delta E^{(1)}\left|\frac{t_*}{t}\right|^{1-\a_0},\\
\Delta E^{(1)} &=& \left[5\ln\frac{m k_0(2,1)}{m_r\,\a_\textsc{qed}^2k_0(2,0)}+\frac{7}{6}+\frac{5m}{8m_r}\right]\nonumber\\
&&\times\frac{\a_\textsc{qed}^5 m}{6\pi}\left(\frac{m_r}{m}\right)^3\nonumber\\
&\simeq& \left[5\ln\frac{k_0(2,1)}{\a_\textsc{qed}^2k_0(2,0)}+\frac{43}{24}\right]\frac{\a_\textsc{qed}^5 m}{6\pi}\,,
\ea
where $\Delta E^{(0)}$ is the standard theoretical prediction, $\Delta E^{(1)}$ is the correction due to anomalous-geometry effects and in the last step we further approximated $m_r/m\approx (0.5107\,{\rm MeV})/(0.5110\,{\rm MeV})\approx 1$. For the hydrogen atom, $k_0(2,0) = 16.64$ and $k(2,1)=0.97$ \cite{DrSw} (reported also in \cite{EGS}), while $\a_\textsc{qed}(t_\textsc{qed})=7.3\times 10^{-3}$ as measured in quantum-electrodynamics experiments. This gives $\Delta E^{(1)}\approx 2\times 10^{-5}\,{\rm eV}$. If we assume that the experimental uncertainty $\delta E\approx 10^{-11}\,{\rm eV}$ in \Eq{hyd} gives an upper bound on the multiscale correction, we can derive an upper bound for the characteristic time $t_*$:
\be\label{upba}
t_* < t_\textsc{qed}\left|\frac{\delta E}{\Delta E^{(1)}}\right|^{\frac{1}{1-\alpha_0}}.
\ee
Taking the upper limit $t_\textsc{qed}=10^{-16}\,{\rm s}$ to be conservative, one can plot the right-hand side of \Eq{upba} as a function of $0<\a_0<1$. We find a global maximum at $\a_0=0$, which yields the absolute upper bound
\be\label{abswe}
t_* < 5\times 10^{-23}\,{\rm s}\,,
\ee
while for $\a_0=1/2$
\be\label{a12we}
t_*^{(\a_0=1/2)} < 2\times 10^{-29}\,{\rm s}\,.
\ee
These are the bounds \Eq{tbou1} and \Eq{tbou3} announced in Sec.\ \ref{goa}.


\section{Standard Model with \texorpdfstring{$q$}{}-derivatives}\label{stmoq}


\subsection{Multiscale Standard Model}\label{qumu}

Contrary to the case with weighted derivatives, theories with $q$-derivatives on multiscale Minkowski spacetime are defined to be invariant under the $q$-Poincaré transformations \Eq{qlort} (and, for the oddness of the geometric coordinates under reflections, also under CPT). The dynamics is therefore straightforward: it is the usual one with the replacement \Eq{xq} and
\be\label{repl}
\cD_\mu\to\p_{q^\m}\,.
\ee
For instance, the Yang--Mills Lagrangian \Eq{lym} is now defined with
\ba
F^a_{\m\n} &=& \frac{\p A^a_\n}{\p q^\mu(x^\mu)} -\frac{\p A^a_\m}{\p q^\nu(x^\nu)} -g'\epsilon^a_{\ bc} A_\m^b A_\n^c\,,\label{fwei2}\\
B_{\m\n} &=&\frac{\p B_\n}{\p q^\mu(x^\mu)} -\frac{\p B_\m}{\p q^\nu(x^\nu)}\,,\label{bwei2}
\ea
instead of Eqs.\ \Eq{fwei} and \Eq{bwei}. All the couplings are constant:
\ba
&&\l=\l_0={\rm const},\qquad w=w_0={\rm const},\\
&& g=g_0={\rm const},\qquad g'=g_0'={\rm const}.
\ea
In the covariant derivatives \Eq{nab1}, \Eq{covL} and \Eq{covL2} one makes the replacement \Eq{repl}. The Lagrangian \Eq{yuka} has a constant Yukawa coupling $G_e$.

Also the sector of strong interactions follows through: the Lagrangian $\cL_{\rm q}+{\cal L}_{\rm Yuk}$ is given by Eqs.\ \Eq{lquarks} and \Eq{LYuk} with the replacement \Eq{repl} in the covariant derivatives and with constant Yukawa couplings
\be
y'=y'_0={\rm const},\qquad y''=y''_0={\rm const}\,.
\ee


\subsection{Physics of the theory with \texorpdfstring{$q$}{}-derivatives}\label{physq}

Now we come to the physical implications of the multiscale theory with $q$-derivatives. Since the frame where physical measurements are performed is established uniquely, it is possible to predict a deviation of particle-physics observables from the standard lore. However, when the action is written explicitly in $x$ coordinates, it resembles an inhomogeneous field theory in ordinary spacetime with noncanonical kinetic terms and nonconstant couplings. For example, the action of a real scalar field with polynomial potential in $1+1$ dimensions would be
\ba
S_\phi&=&\int\rmd^2 q\,\left\{\frac12[\p_{q_0(t)}\phi]^2-\frac12[\p_{q_1(x)}\phi]^2-\sum_n\l_n\phi^n\right\}\nonumber\\
&=&\int\rmd^2 x\,\left\{\vphantom{\sum_n}\frac{v_1(x)}{2v_0(t)}\dot\phi^2-\frac{v_0(t)}{2v_1(x)}(\p_x\phi)^2\right.\nonumber\\
&&\qquad\qquad\left.-\sum_n[v_0(t)v_1(x)\l_n]\phi^n\right\},
\ea
where we have ignored gravity. From this point on, we proceed as in the case with weighted derivatives. Since we do not know how to define a quantum field theory with varying couplings and nonhomogeneous kinetic terms, it is necessary to perform all calculations in geometric coordinates. Therefore, we transform to the integer picture  via \Eq{xq} where the theory looks trivial and one can borrow all the known calculations in the Standard Model. At the end of the day, any ``time'' or ``spatial'' interval or ``energy'' predicted are not a physical time or spatial interval or energy, since they are measured with $q$-clocks, $q$-rods or $q$-detectors. The results must be reconverted to the fractional picture to interpret them correctly. A discussion on the use of nonadaptive clocks and rods at different scales can be found in \cite{fra7}.

In the next subsections, we illustrate the idea with the examples of the muon decay rate and of the Lamb shift.


\subsection{Muon decay rate}\label{sec:feynman}

Consider a massive particle with mass $m$ in ordinary spacetime. Its quantum propagator in momentum space is proportional to $[k^2+m^2+\Pi(k^2)]^{-1}$; at one loop, $\Pi(k^2)$ is the contribution of the one-particle irreducible bubble diagrams. In the on-shell regularization scheme, $m^2$ is the physical mass and the propagator has a simple pole at $k^2=-m^2$, so that $\Pi(-m^2)=0$. Calculating $\Pi(-m^2)$ and imposing that it vanishes determines the counterterm to be added to the Lagrangian. However, if $\Pi(-m^2)=:\rmi m\Gamma$ is purely imaginary one is meeting a resonance, i.e., an unstable particle. In this case, in a neighborhood of the mass shell, the propagator can be written as $\propto (k^2+m^2+\rmi m \Gamma)^{-1}$, where $\Gamma$ is called decay width and has the dimension of a mass. The name stems from the fact that the propagator in the rest frame is proportional to the quantum amplitude describing the decay of the resonance. The square of the amplitude is the relativistic Breit--Wigner probability distribution $f_{\rm BW}(E)=c(m,\Gamma)\,\Gamma/[(m^2-E^2)^2+ (m\Gamma)^2]$, where $E$ is the resonance energy in the center of mass and $c$ is a constant whose dependence on $\Gamma$ is such that $c\to 2m^2/\pi$ and $f_{\rm BW}(E)\to 2m\delta(E^2-m^2)$ in the limit $\Gamma\to 0$; this distribution is sharply peaked at $E=m$.

The decay width can be calculated explicitly for the unstable particles appearing in the Standard Model. To a scattering process described by a one-particle initial state $|{\rm i}\rangle$ and a many-particle final state $|{\rm f}\rangle$, one associates the Feynman amplitude $\langle {\rm f}\vert {\rm i}\rangle$ which is computed according to the particles involved and up to a certain perturbative order. From the (nonnormalized) transition probability $\cP({\rm i}\ra {\rm f}) = |\langle {\rm f}\vert {\rm i}\rangle|^2$, one extracts the decay rate $\Gamma$ for the resonance $|{\rm i}\rangle$. In the case of the muon, the process is $\mu^- \ra e^- \ove{\nu}_e \nu_{\mu}$ and it is mediated by a gauge boson $W$. Neglecting the masses of the electron $e^-$ and the neutrino $\nu_e$, in $D=4$ one has
\be \label{gamma0}
\Gamma = \frac{G^2_{\rm F} m_{\rm mu}^5}{192\pi^3}+\cdots\,,
\ee
where $G_{\rm F}=\sqrt{2}g_0^2/(8M_{\rm W}^2)$ is Fermi constant, $m_{\rm mu}$ is the muon mass and the ellipsis denotes loop corrections to the tree-level contribution. The mean lifetime of the muon is defined by
\be\label{restim}
\tau_{\rm mu}=\tau_0:=\frac{\hbar}{\Gamma}\qquad \text{(in ordinary spacetime)}.
\ee

Let us now see the case of multiscale spacetimes. In the theory with weighted derivatives, the propagator is the same as the usual one up to a measure-dependent normalization \cite{frc6} and the decay rate $\Gamma$ is defined exactly in the same way.\footnote{For models with weighted derivatives, quantum-mechanical and stochastic probability distributions usually differ from the standard ones only by an energy-dependent normalization \cite{frc7,frc14}. This normalization can actually change the profile of the density of states, since it is measure dependent and it can be singular at the special points of the measure. Therefore, in the fractional picture the Breit--Wigner distribution would be something of the form $C(E)\, f_{\rm BW}(E)$ and it would not be possible to interpret $\Gamma$ as the width. However, on one hand there does not seem to be easy alternative ways to define the decay width in the fractional picture (after all, $\Gamma$ is introduced from the propagator and the latter is trivially modified \cite{frc6}) and, on the other hand, the interpretation of $\Gamma$ is clear in the integer picture and does not require such modifications.} The quantum field theory is dealt with in the integer picture, the final tree-level result is \Eq{gamma0} (clearly, also loop corrections would follow through the standard calculation), there are no unit changes when reverting back to the fractional picture and the mean lifetime of the muon is \Eq{restim}: the physics is insensitive to the anomalous properties of the geometry.

In the theory with $q$-derivatives, one works in the integer picture and obtains \Eq{gamma0}. However, $\Gamma$ is no longer the inverse of the muon lifetime. The propagator of the resonance is $\propto [p(k)^2+m_{\rm mu}^2+\rmi m_{\rm mu}\Gamma]^{-1}$ and $\Gamma$ is still the width of the Breit--Wigner distribution, but the inverse of $\Gamma$ is a composite object. From the form of the propagator, it is natural to make the identification
\be
\Gamma=p^0\left(\frac{\hbar}{\tau_{\rm mu}}\right)\stackrel{\text{\tiny \Eq{pik}}}{=}\frac{1}{q^0(\tau_{\rm mu}/\hbar)}\,,
\ee
and the physically observed muon lifetime is found by inverting the relation (from now on, $\hbar=1$)
\be\label{q0mu}
q^0(\tau_{\rm mu})=\frac{1}{\Gamma}=\tau_0\qquad \text{(in multiscale spacetime)}.
\ee
The replacement of \Eq{restim} with formula \Eq{q0mu}, valid in multiscale spacetimes with $q$-derivatives, gives a characteristic prediction that can be compared with that in standard spacetime. For practical purposes, a constraint on the fundamental scales in the measure can be obtained as follows. First, we make a choice of geometric measure. The binomial measure \Eq{binomialm2} is enough to extract interesting information:
\be\label{ansq1}
q_*(\tau_{\rm mu}) = \tau_{\rm mu}+\frac{t_*}{\a_0}\left(\frac{\tau_{\rm mu}}{t_*}\right)^{\a_0}=\tau_0\,.
\ee
The muon lifetime is not observed directly. Experiments determine the Fermi constant $G_{\rm F}= 1.1663787(6)\times 10^{-5}\,\text{GeV}^{-2}$ and the muon mass $m_{\rm mu} = 105.6583715(35) \,\text{MeV}$ \cite{pdg}. Using \Eq{gamma0}, one has \cite{pdg}
\be\label{1gaex}
\tau_0=2.1969811(22)\times 10^{-6}\,{\rm s}
\ee
for $\mu^-$. The lifetime of $\mu^+$ is almost the same and we can ignore the difference. If we knew both $\a_0$ and $t_*$, we would invert \Eq{ansq1} and find the multiscale prediction for $\tau_{\rm mu}$. As we do not, we opt for a different approach. We assume realistically that $t_*$ is small enough so that the scale-dependent part of the measure is small and $\tau_{\rm mu}\approx \tau_0$ to a very good approximation. Then, we account for all the experimental error $\delta\tau\approx 6.6\times 10^{-12}\,{\rm s}$ at the $3\s$-level as setting an upper limit on the effects of anomalous geometry:
\be\nonumber
\frac{t_*}{\a_0}\left(\frac{\tau_0}{t_*}\right)^{\a_0}<\delta\tau\,,
\ee
implying that
\be\label{tstfin}
t_*< \left(\frac{\a_0\delta\tau}{\tau_0^{\a_0}}\right)^{\frac{1}{1-\a_0}}\,.
\ee
Computing \Eq{tstfin} as a function of $0<\a_0<1$, we find that the maximum $t_*$ is attained for $\a_0\approx 0.06$. This value of $\a_0$ has no special meaning in the theory but it sets the absolute upper bound in \Eq{tbou}. On the other hand, for the central value $\a_0=1/2$ (which can have some theoretical justification \cite{frc1,frc2}) the allowed range $t<t_{\rm max}^{(\a_0=1/2)}$ is lowered by 5 orders of magnitude.\footnote{Due to some freedom in the normalization of the factor $(t/t_*)^{\a_0}$, one can slightly change the above bounds but not by much. Replacing $\a_0\to \Gamma(1+\a_0)$ in the numerator of \Eq{tstfin} (as in the original definition of fractional measures, where $\Gamma$ is Euler's function), Eq.\ \Eq{tbou} becomes $t_* < 10^{-12}\,{\rm s}$ (at $\a_0=0$) and $t_*^{(\alpha_0=1/2)} < 10^{-18}\,{\rm s}$.}


\subsection{Lamb shift}\label{sec:lamb}

Independent bounds on the scales of the theory come from quantum electrodynamics and the Lamb shift effect. Following a procedure analogous to the one for the muon lifetime, we use the experimental uncertainty to determine the fundamental energy $E_*$ below which the effects of the anomalous geometry become negligible. The theoretical calculation of the radiative corrections to the Lamb shift is identical to the ordinary one upon the replacement $E\to p^0(E)$ according to the momentum geometric coordinates \Eq{pik}. Since we expect $E_*$ to be much larger than the characteristic energy scale involved in these experiments, we can make the approximation $E_*\gg E$ in \Eq{bino2}. A check \emph{a posteriori} will confirm this step. Considering the binomial measure for $0<\a_0<1$, one has
\be \label{peb}
p_*(E) \simeq E-\frac{|E|}{\alpha_0}\left|\frac{E_*}{E}\right|^{\alpha_0-1}, 
\ee
so that the difference $\Delta p_*(E)=p_*(E_1)-p_*(E_2)$ between geometric energies is related to the difference $\Delta E=E_1-E_2$ between energies by 
\ba
\Delta p_*(E) &\simeq& \Delta E+\frac{E_*^{\a_0-1}}{\a_0}\left(|E_2|^{2-\a_0}-|E_1|^{2-\a_0}\right)\nonumber\\
&\simeq& \Delta E+\frac{2-\a_0}{\a_0}\left|\frac{E_1}{E_*}\right|^{1-\a_0}(|E_2|-|E_1|),\nonumber
\ea
where in the second line we have used the fact that, for the levels $2S$ and $2P$ of hydrogenic atoms, $\Delta E/E_1\sim \Delta E/E_2\ll 1$. The expansion $x^a-1=a(x-1)+O[(x-1)^2]$ then applies. Letting $D=4$, identifying $E_1=E_{2S}$ and $E_2=E_{2P}$ with the energy of, respectively, the $2S_{1/2}$ and $2P_{1/2}$ state and noting that both $E_{2S}$ and $E_{2P}$ are negative, the relation between geometric and physical Lamb shift is
\be\label{pfin}
\Delta p_*(E) \simeq \Delta E+\frac{2-\a_0}{\a_0} \Delta E \left|\frac{E_{2S}}{E_*}\right|^{1-\a_0}.
\ee
Since the multiscale correction is going to be small, it is safe to assume that $\Delta p_*(E)\simeq \Delta E$. Then, the second term in \Eq{pfin} cannot be larger than the experimental error $\delta E$, which establishes a lower bound for the energy $E_*$:
\be\label{pfin2}
E_* >\left(\frac{\a_0}{2-\a_0}\frac{\delta E}{\Delta E}\right)^{\frac{1}{\a_0-1}}|E_{2S}|\,.
\ee
The smaller the experimental uncertainty $\delta E/\Delta E$ and the energies $|E_{1,2}|$ involved, the larger the lower bound on $E_*$. From Eq.\ \Eq{hyd}, the relative experimental uncertainty on the $2S$-$2P$ Lamb shift of hydrogen is $\delta E/\Delta E\approx 2.8\times 10^{-6}$ at $1\s$ confidence level, the same as for deuterium (for helium, $\delta E/\Delta E\approx 1.5\times 10^{-5}$). Rounding up to the $3\s$ level, 
\be
\frac{\delta E}{\Delta E}\approx 8.2\times 10^{-6}\,.
\ee
The energy of the $2S_{1/2}$ state is $E_{2S}\approx-3.4\,\text{eV}$. Plugging these values into \Eq{pfin2}, the right-hand side has a minimum at (again) $\a_0\approx 0.06$, resulting in the absolute lower bound in \Eq{Elow}. Consistently, $|E_{2S}|/E_*\ll 1$. For the preferred value $\a_0=1/2$, the lower bound is much larger, $E_*>E_{\rm min}^{(\a_0=1/2)}=450\,\text{GeV}$.


\section{Discussion}\label{sec:conc}

When the dimension of spacetime changes by virtue of exotic physics independent of curvature corrections, the dynamics of quantum particles is usually affected. Modified dispersion relations, quantum geometries and multifractal backgrounds are all characterized by dimensional flow in one way or another. However, in this paper we have shown that the case of multiscale spacetimes with weighted derivatives is special inasmuch as the observables of quantum field theory with non-Abelian gauge fields are insensitive to dimensional flow: the latter cannot be tested in accelerator experiments in a world described by such model. Only in the $U(1)$ sector (electrodynamics) or when gravity is turned on does the scale hierarchy of the geometry manifest itself, in such a way that the dynamics is fundamentally different from more traditional settings such as scalar-tensor theories \cite{frc11}. This result formally concludes the basic formulation of the theory with weighted derivatives and suggests that future investigation be carried out mainly in the context of astrophysics and cosmology. On the other hand, the multiscale theory with $q$-derivatives is nontrivial in all gauge sectors.

Particle-physics observations can place bounds on the characteristic scales of the geometry of both theories. The method we employed to extract this information is crude but effective, as also shown in early applications to dimensional-regularization toy models \cite{ScM,ZS,MuS}: the $3\s$-level experimental uncertainty is used as an upper bound on possible multiscale effects. We extracted the absolute and characteristic upper bounds \Eq{tbou} on the time scale $t_*$ in the hierarchy by comparing the muon lifetime predicted by the theory with experiments. Similarly, the upper bounds \Eq{tbou1}--\Eq{tbou3} and the lower bounds \Eq{Elow} were obtained from the $2S$-$2P$ Lamb shift effect in hydrogenlike atoms in the theory with, respectively, weighted and $q$-derivatives. All these bounds are more sophisticated than those found in the toy models mentioned above, which are not multiscale: in general, a spacetime with a fixed dimension $D$ different from 4 gives rise to much less flexible phenomenology, which invariably ends up in scale-independent constraints $|D-4|\sim 10^{-5}\mhyp 10^{-11}$ (see \cite{frc2} for a summary of these old results).

Further aspects that deserve consideration for future investigations are related to the quantization of multiscale Standard Models. In the present paper, when quantum corrections were considered, the result was simply imported from the corresponding standard result in the integer picture. However, we have not discussed the fate of quantum symmetries, which are expected to happen in the presence of quantum-gravity effects and might be broken also in the case of multiscale theories in the fractional picture. Violations of global symmetries such as CPT (which are preserved classically, as we have seen) can lead to a baryon-number violation signaled by  proton decays or neutron-antineutron transitions accessible in the laboratory. It would be interesting to find and compare these limits with those discussed above.

On the theoretical side, two other questions are how the exotic measure of multiscale geometry enters the effective action and how it affects the path integral. Concerning the first point, a new type of nonlocality in the momentum should arise as a consequence of having deformed the propagator as described in \cite{frc6} and here. The second point is perhaps subtler if one thinks that $\hbar$ enters the path integral just as a rescaling
\be
v(x) \to \frac{v(x)}{\hbar}\,,
\ee
so that the effect of the measure weight decorating the action integral may be seen as a spacetime-dependent Planck constant. This suggests the idea that the classical limit $\hbar \to 0$ can be modulated by a specific regime of the multiscale measure. To be more specific, one should carry out a study along the lines of \cite{frc5}, where the multifractional path integral for a quantum-mechanical particle was constructed.

We conclude by relating the time and energy bounds found in this work. So far, we have followed a conservative approach and treated the fundamental length, time and energy scales $\ell_*$, $t_*$ and $E_*$ in the binomial measure as independent. A drastic simplification of the theory would occur if all these scales were related to one another by some unit conversion. In a standard setting, one would make such conversion using Planck units. Here, the most fundamental scale of the system is the one appearing in the full measure with logarithmic oscillations \cite{fra4,frc2}, denoted as $\ell_\infty$ in Eq.\ \Eq{fom}. For the time direction one has a scale $t_\infty$, while in the measure in momentum space the fundamental energy $E_\infty$ would appear. Then, one may postulate that the scales $\ell_*\gg\ell_\infty$, $t_*\gg t_\infty$ and $E_*\ll E_\infty$ are related by
\be\label{formu}
E_*=\frac{t_\infty E_\infty}{t_*}\,,\qquad t_*=\frac{t_\infty \ell_*}{\ell_\infty}\,.
\ee
The origin of these formul\ae\ is mysterious: in their present formulation, theories of multiscale spacetimes do not require this mutual dependence of the scales in the hierarchy. Nevertheless, it is intriguing to explore the consequences of \Eq{formu}. We recall that log-oscillating measures provide an elegant extension of noncommutative $\kappa$-Minkowski spacetime and explain why the Planck scale does not appear in the effective measure thereon \cite{ACOS}. In turn, this connection suggests that the fundamental scales in the log oscillations coincide with the Planck scales:
\be\label{infpl}
t_\infty=t_\Pl\,,\qquad \ell_\infty=\ell_\Pl\,,\qquad E_\infty=m_\Pl\,.
\ee
In four dimensions, $t_\Pl =\sqrt{\hbar G/c^3}\approx 5.3912 \times 10^{-44}\,\text{s}$, $\ell_\Pl=\sqrt{\hbar G/c^5} \approx 1.6163 \times 10^{-35}\,\text{m}$ and $m_\Pl=\sqrt{\hbar c/G} \approx 1.2209\times 10^{19}\,\text{GeV}c^{-2}$. Remarkably, Eq.\ \Eq{infpl} connects, via Newton's constant, the prefixed multiscale structure of the measure with the otherwise independent dynamical part of the geometry. Also, it makes the log-oscillating part of multiscale measures [and so the whole measure, via \Eq{formu}] intrinsically quantum in the sense that Planck's constant $\hbar=h/(2\pi)$ appears in the geometry.\footnote{Moreover, putting time and space on an equal footing as in \Eq{formu} may indicate a similar symmetry for geometric coordinates at all scales, which implies isotropy of the fractional indices: $\a_\mu=\a$ for all $\mu=0,1,\dots,D-1$. This further assumption is not necessary for our analysis.}

In the light of Eqs.\ \Eq{formu} and \Eq{infpl}, we can manipulate the bounds $t_*<t_{\rm max}$ and $E_*>E_{\rm min}$ we have obtained on $t_*$ and $E_*$ to extract new bounds summarized in Tables \ref{tab1} and \ref{tab2}. For each part of the tables (absolute bounds and bounds with $\a_0=1/2$), the ``muon lifetime'' row is $(t_{\rm max},\,\ell_{\max}= t_{\rm max} \ell_\Pl/t_\Pl,\,\bar E_{\rm min}= m_\Pl t_\Pl/t_{\rm max})$, while the ``Lamb shift'' row is $(\bar t_{\rm max}=t_\Pl m_\Pl/E_{\rm min},\,\bar\ell_{\max}= \ell_\Pl m_\Pl/E_{\rm min},\, E_{\rm min})$.
\begin{table}[ht]
\begin{center}
\caption{Absolute and preferred ($\a_0=1/2$) bounds on the hierarchy of multiscale spacetimes with weighted derivatives. The bounds \Eq{abswe} and \Eq{a12we} obtained directly from experiments, without the input \Eq{formu}--\Eq{infpl}, are in boldface. All figures are rounded.\label{tab1}}
\begin{tabular*}{\columnwidth}{l @{\extracolsep{\fill}} ccc}\hline\hline
Absolute bounds & $t_*$ (s)         & $\ell_*$ (m) & $E_*$ (eV) \\\hline
Lamb shift      & ${\bf <10^{-23}}$       & $<10^{-14}$  & $>10^{7}$ \\\hline\hline
$\a_0=1/2$      & $t_*$ (s)         & $\ell_*$ (m) & $E_*$ (eV) \\\hline
Lamb shift      & ${\bf <10^{-29}}$       & $<10^{-20}$  & $>10^{13}$ \\\hline\hline
\end{tabular*}
\end{center}
\end{table}
\begin{table}[ht]
\begin{center}
\caption{Absolute and preferred ($\a_0=1/2$) bounds on the hierarchy of multiscale spacetimes with $q$-derivatives. Bounds obtained directly from experiments are in boldface. All figures are rounded.\label{tab2}}
\begin{tabular*}{\columnwidth}{l @{\extracolsep{\fill}} ccc}\hline\hline
Absolute bounds & $t_*$ (s)         & $\ell_*$ (m) & $E_*$ (eV) \\\hline
muon lifetime   & ${\bf <10^{-13}}$ & $<10^{-5}$   & $> 10^{-3}$ \\
Lamb shift      & $<10^{-23}$       & $<10^{-15}$  & ${\bf >10^7}$ \\\hline\hline
$\a_0=1/2$      & $t_*$ (s)         & $\ell_*$ (m) & $E_*$ (eV) \\\hline
muon lifetime   & ${\bf <10^{-18}}$ & $<10^{-9}$   & $> 10^{2}$ \\
Lamb shift      & $<10^{-27}$       & $<10^{-19}$  & ${\bf >10^{11}}$ \\\hline\hline
\end{tabular*}
\end{center}
\end{table}
In the theory with $q$-derivatives, the bounds from the Lamb shift are much stronger than those from the decay rate of the muon. The characteristic scales $\ell_*$ and $t_*$ cannot be larger than about $10^{17}$ times the Planck scale. In particular, the $\a_0=1/2$ bound on $\ell_*$ is stronger than the heuristic estimate $\ell_*<10^{-18}\,{\rm m}$ made in \cite{frc2}. For the Lamb shift, the bounds in the theory with weighted derivatives are one or two orders of magnitude stronger than those for the theory with $q$-derivatives. Interestingly, the $\a_0=1/2$ Lamb-shift bounds $E_*>28\,{\rm TeV}$ (weighted derivatives) and $E_*>450\,\text{GeV}$ ($q$-derivatives) are not far from the energies currently probed in the LHC.

Multiscale theories are not the only context where characteristic scales appear and can be constrained. Just to give one example, in string field theory certain nonlocal operators modify the physics at scales close to the string length $\sqrt{\a'}=l_{\rm s}$. This scale is supposed to be extremely small but it can be constrained by experiments without any theoretical prejudice on its size. LHC data bound this scale as $l_{\rm s}< 10^{-19}\,\text{m}$, corresponding to $E_{\rm s}> 10^3\,\text{GeV}$ \cite{BiOk}, while observations on optomechanical heavy quantum oscillators give $l_{\rm s}< 10^{-15}\,\text{m}$ \cite{Bel15}. These figures are similar to those we found in this paper.

Let us also compare the numbers in the tables with the characteristic length and time scales of particle interactions. The electromagnetic force propagates indefinitely, so that $\ell_\textsc{qed}=\infty$; on the other hand, $t_\textsc{qed}\sim 10^{-21}-10^{-16}\,{\rm s}$. The length scale of the weak interaction is $\ell_{\rm weak}=\hbar/(m_W c)\sim 10^{-18}\,{\rm m}$, while $t_{\rm weak}> 10^{-10}\,{\rm s}$. For the strong interactions, $\ell_\textsc{qcd}\simeq\hbar/(m_\pi c)\sim 10^{-15}\,{\rm m}$ (where $\pi$ is the lightest massive meson) and $t_\textsc{qcd}\simeq \ell_\textsc{qcd}/c\sim 10^{-23}\,{\rm s}$. Since the preferred upper bounds on $t_*$ and $\ell_*$ coming from the Lamb shift are smaller than all these characteristic scales, it is reasonable to conclude that, for all practical purposes, electroweak and strong interactions cannot feel multiscale effects in any of the theories considered here.

We have stressed at several points in the paper that the physics is described by the fractional picture, while the integer picture is just an auxiliary tool. Contrary to a similar dilemma in scalar-tensor theories, here we do not have a choice in selecting the frame where observables should be extracted, even at the classical level. Therefore, the violation of Lorentz invariance in the fractional picture is another element we could constrain multifractal geometry with. Effects of Lorentz-symmetry breaking in the Standard Model have repercussions not only in accelerator experiments but also in cosmic-ray and neutrino physics \cite{CoGl,KoMe}. Constraints from these observations are typically more stringent than those found in this paper from the muon decay rate and the Lamb shift and they may provide a very useful source of information to further pin down the range of validity of multiscale theories. Data from yet other physical systems or processes can provide a cross-check of the above bounds. In particular, one could study the multiscale version of massive quantum oscillators and use those observations to get independent constraints, especially because table-top high-precision experiments already under construction will be able to improve such bounds by several orders of magnitude \cite{Bel15}. We also foresee a number of cosmological applications, some of which have already been worked out \cite{frc11} or will appear soon \cite{frc14}.


\section*{Acknowledgments} 

The work of G.C.\ is under a Ram\'on y Cajal contract and is supported by the I+D grant FIS2014-54800-C2-2-P. D.R.F.\ is supported by the GRUPIN 14-108 research grant from Principado de Asturias. D.R.F.\ is thankful to Alessandro Pini for useful discussions. 

\medskip

\paragraph*{Note added.---}After the submission of this paper to the electronic preprint archive, it was suggested to us to look at the anomalous magnetic moment of the electron, the $g-2$ factor, as a possible source of stronger constraints on multiscale geometry. A short calculation gives an interesting result for the theory with weighted derivatives. From the triangular vertex in the integer picture, at one loop it is known that $g-2=\tilde\a_\textsc{qed}/\pi$. From now on, we consider just the fine-structure constant, which is measured with an accuracy of $\delta\a_\textsc{qed}/\a_\textsc{qed}\sim 10^{-10}$. Since the measured fine-structure constant is $\a_\textsc{qed}(t)=\tilde\a_\textsc{qed}/v_*(t)$, for the binomial measure \Eq{binomialm} the difference between the integer and fractional constant is $\Delta\a_\textsc{qed}=\a_\textsc{qed}(t)|t_*/t|^{1-\a_0}$. Demanding $\Delta\a_\textsc{qed}<\delta\a_\textsc{qed}$ and setting $t=t_\textsc{qed}=10^{-16}\,{\rm s}$, we obtain $t_*< 10^{-16-10/(1-\a_0)}\,{\rm s}$. Therefore, we get an upper bound on $t_*$ at $\a_0=0$, generating also length and energy bounds:
\be
t_*<10^{-26}\,{\rm s}\,,\quad \ell_*<10^{-17}\,{\rm m}\,,\quad E_*>10\,{\rm GeV}\,,
\ee
which are three orders of magnitude stronger than the Lamb-shift absolute bounds in Table \ref{tab1}. For $\a_0=1/2$, we get
\ba
&&t_*^{(\a_0=1/2)}<10^{-36}\,{\rm s}\,,\nonumber\\
&&\ell_*^{(\a_0=1/2)}<10^{-27}\,{\rm m}\,,\\
&&E_*^{(\a_0=1/2)}>10^{11}\,{\rm GeV}\,,\nonumber
\ea
seven orders of magnitude stronger than the Lamb-shift $\a_0=1/2$ bounds in Table \ref{tab1}.

The theory with $q$-derivatives is immune to similar constraints because it predicts the same $g-2$ factor and fine-structure constant as in the ordinary Standard Model. It is easy to understand why. As we have seen, the way the $q$-theory conveys multiscale effects to physical observables is via a transition from adaptive measurement units (integer picture) to nonadaptive ones (fractional picture). In the case of the Lamb shift, we borrowed the standard QED result for the shift in the energy levels and applied it to the difference $\Delta p_*(E)$ between geometric energies; then, from $\Delta p_*(E)$ we extracted the actual Lamb shift $\Delta E$ and proceeded with the comparison with experiments. We could have done essentially the same thing by looking at the hydrogen spectrum on a photographic plate. If the theory in the integer picture predicts that two lines A and B are separated by $\Delta X$ microns, then we must interpret $\Delta X=\Delta q_*(x)$ as a composite distance, from which one can calculate the distance $\Delta x$ measured by our physical rulers. And so on. However, \emph{dimensionless} quantities such as $\a_\textsc{qed}$ and $g-2$ are unaffected by having worked with composite momentum or position coordinates. Therefore, these fundamental\footnote{By fundamental, we mean that they are not obtained from the composition of other directly measurable quantities; see \cite{turtl} for an example of nonfundamental observables that can discriminate the theory.} dimensionless observables remain the same in the $q$-theory. Curiously, this situation is complementary to the one for the muon lifetime, where the $q$-theory was sensitive to changes in the geometry while the theory with weighted derivative was not.

\end{document}